\documentclass{aa}  
\usepackage{graphicx}
\usepackage{eqnarray,amsmath}
\usepackage{multicol}
\usepackage[hypcap]{caption}
\usepackage{txfonts}
\usepackage{hyperref}
\usepackage{xcolor}
\graphicspath{{./figures/}} 
\begin{document} 
\titlerunning{Evolution of massive stars with new hydrodynamic wind models}
\title{Evolution of massive stars with new hydrodynamic wind models}
\authorrunning{Gormaz-Matamala et al.}
\author{A. C. Gormaz-Matamala\inst{1,2,3}
\and
M. Curé\inst{2,4}
\and
G. Meynet\inst{5}
\and
J. Cuadra\inst{1}
\and
J. H. Groh\inst{6}
\and
L. J. Murphy\inst{6}}
\institute{Departamento de Ciencias, Facultad de Artes Liberales, Universidad Adolfo Ib\'a\~nez, Av. Padre Hurtado 750, Vi\~na del Mar, Chile\\
\email{alex.gormaz@uv.cl}
\and
Instituto de Física y Astronomía, Universidad de Valparaíso. Av. Gran Breta\~na 1111, Casilla 5030, Valpara\'iso, Chile.
\and
Instituto de Astrofísica, Facultad de Física, Pontificia Universidad Católica de Chile, 782-0436 Santiago, Chile
\and
Centro de Astrofísica, Universidad de Valparaíso. Av. Gran Breta\~na 1111, Casilla 5030, Valpara\'io, Chile.
\and
Geneva Observatory, University of Geneva, Maillettes 51, 1290 Sauverny, Switzerland
\and
School of Physics, Trinity College Dublin, University of Dublin, Dublin, Ireland.}

\date{}

\abstract 
{Mass-loss by radiatively line-driven winds is central to our understanding of massive star evolution either single or in multiple systems.
It for instance plays a key role in making massive star evolution different at different metallicities, specially in the case of very massive stars ($M_*\ge25M_\odot$).}
{Here we present evolutionary models for a set of massive stars, introducing a new prescription for the mass-loss rate obtained from hydrodynamical calculations in which the wind velocity profile, $\varv(r)$, and the line-acceleration, $g_\text{line}$, are obtained in a self consistently way.
These new prescriptions cover the most part of the Main-Sequence phase of O-type stars.}
{We perform a grid of self-consistent mass-loss rates $\dot M_\text{sc}$, for a set of standard evolutionary tracks (i.e., using the old prescription for mass-loss rate) under different values for initial mass and metallicity.
Based in this grid, we elaborate a statistical analysis to create a new simple formula predicting the values of $\dot M_\text{sc}$ just from the stellar parameters, without assuming any extra condition for the wind description.
Therefore, replacing mass-loss rates at the Main Sequence stage from the standard Vink's formula by our new recipe, we generate a new set of evolutionary tracks for $M_\text{ZAMS}=25,40,70$ and $120\,M_\odot$ and metallicities $Z=0.014$ (Galactic), $Z=0.006$ (LMC), and $Z=0.002$ (SMC).}
{Our new derived formula for mass-loss rate predicts a dependence $\dot M\propto Z^a$, where $a$ is not longer constant but dependent on the stellar mass: ranging from $a\sim0.53$ when $M_*\sim120\;M_\odot$, to $a\sim1.02$ when $M_*\sim25\;M_\odot$.

We found important differences between standard and our new self-consistent tracks.
Models adopting the new recipe for $\dot M$ (which starts being around $\sim3$ times weaker than mass-loss rate from the old formulation) retain more mass during their evolution, which is expressed in larger radii and consequently more luminous tracks over the Hertzsprung-Russell diagram.
These differences are more prominent for the cases of $M_\text{ZAMS}=70$ and 120 $M_\odot$ at solar metallicity, where we found self-consistent tracks are $\sim0.1$ dex brighter and keep extra mass up to 20 $M_\odot$, compared with the classical models using the previous formulation for mass-loss rate.
Later increments in the mass-loss rate for tracks when self-consistency is not longer used, attributed to the LBV stage, produce different final stellar radii and masses before the end of H-burning stage, which are analysed case to case.

Moreover, we observed remarkable differences for the evolution of the radionuclide isotope $^{26}$Al in the core and the surface of the star.
Since $\dot M_\text{sc}$ are weaker than the commonly adopted values for evolutionary tracks, self-consistent tracks predict a later modification in the abundance number of $^{26}$Al in the stellar winds.
This new behaviour could provide useful information about the real contribution of this isotope from massive stars to the Galactic interstellar medium.}
{}

\keywords{Hydrodynamics -- Stars: early-type -- Stars: evolution -- Stars: winds, outflows -- Stars: mass-loss}
\maketitle

\section{Introduction}
	Evolution of massive stars (with $M_*\gtrsim8\,M_\odot$) is an important topic in Stellar Astrophysics because they end their lives with core-collapse events as supernovae, resulting in remnants such as neutron stars or black holes depending on the initial stellar mass \citep[see review from][]{heger03}.
	Subsequently, massive stars are also important for the study of nucleosynthesis, production of ionising flux, feedback due to wind momentum, studies of star formation history and galaxy evolution.
	The main feature that characterises massive stars and their evolution, are their powerful stellar winds.
	Indeed, for the case of very massive stars with $M_*\gtrsim25\,M_\odot$, the amount of matter released by these stellar outflows (namely mass-loss rate, $\dot M$) is critical for the understanding of the evolution of these stars \citep{vink21b}.

	Details of the effects produced by the stellar winds and the mass-loss rate over the evolution of massive stars were outlined by \citet{maeder87b} and more recently by \citet{langer12}, \citet{groh14} and \citet{meynet15}.
	The work from \citet{meynet94} taught us that even changes on $\dot M$ by a factor of two may dramatically affect the fate of a star, whereas more recent studies \citep{brott11,ekstrom12,georgy13,groh19,eggenberger21} have confirmed the same trend.
	Thus, studies constraining the real mass-loss rate of massive stars, either from observations of from theoretical framework, are crucial.
	One of such studies providing theoretical values for the mass-loss rate set by the so-called \textit{Vink's formula} \citep{vink00,vink01}, which have been extensively implemented for the development of multiple evolutionary tracks.
	However, diagnostics of mass-loss rates performed in the recent years consider that values from Vink's formula are overestimated by a factor of $\sim3$ \citep{bouret12,surlan13,vink21b}.
	Hence, more updated prescriptions for $\dot M$ are required, to revisit the evolutionary tracks of massive stars.
	
	In the last decades, there have been large advances towards a more detailed description of the winds of massive stars, and therefore providing better diagnostics for their mass-loss rates.
	The progress made by radiative transfer codes such as FASTWIND \citep{santolaya97,puls05}, CMFGEN \citep{hillier90a,hillier90b,hillier98} or PoWR \citep{grafener02,hamann03}, and their synthetic spectra, contributed with more accurate observational constrains on $\dot M$ based on spectral fittings over diagnostic lines such as H$\alpha$ or P-Cygni profiles in the ultraviolet range.
	These evaluations of mass-loss rate also consider the presence of inhomogeneities (clumping) in the wind, which evidenced an overestimation of the real value of $\dot M$ compared with when the homogeneous models were adopted \citep{bouret05}.
	Since then, clumping has been an important issue to consider for the spectral fitting and the constrain of mass-loss rate, and the treatment of these inhomogeneities have been a matter of study in the last decade \citep{surlan12,surlan13,sundqvist18}.
	
	From the hydrodynamics of the line-driven theory, there have been advances too.
	Different studies have made efforts to calculate mass-loss rate, by solving wind hydrodynamics and including a more coherent description for the radiative acceleration.
	This, in order to provide a more physically self-consistent solution between the wind hydrodynamics and the line-driven process.
	We highlight the Monte-Carlo simulations performed by \citet{muijres12}, whose results show an excellent agreement with \citet{vink00,vink01}.
	However, more recent studies determined values for mass-loss rate more in agreement with the clumped diagnostics, and then below the values for $\dot M$ obtained from Vink's formula.
	Some of these studies calculated the radiative acceleration by solving radiative transfer, either with PoWR \citep{sander17}, FASTWIND \citep{sundqvist19,bjorklund21} or CMFGEN \citep{alex21a}.
	Besides, we have the more versatile frameworks from \citet{kk17,kk18} and \citet{alex19,alex21b}, which provide theoretical self-consistent values of mass-loss rate for a large grid of different stellar conditions.
	Given that these results are the state-of-the-art, studies of the evolution of massive stars establishing a coherence between the line-acceleration and the hydrodynamics of their stellar winds are not fully performed.

	In this work, we present new evolutionary tracks calculated with the Geneva evolutive code \citep[][hereafter \textsc{Genec}]{maeder83,maeder87a}, implementing as recipe for the mass-loss rates the self-consistent m-CAK solutions for the stellar wind performed by \citet{alex19} instead the classical Vink's formula.
	This self-consistent m-CAK prescription has demonstrate to provide us reliable theoretical values for $\dot M$, together with achieving adequate spectral fittings for observations of O-type stars \citet{alex21b}.
	Thus, self-consistent mass-loss rates (hereafter $\dot M_\text{sc}$) can be implemented as an input ingredient for the calculation and study of the evolution of massive stars.
	Differences between the resulting `self-consistent' evolutionary tracks are analysed, in comparison with the `standard' tracks based on the Vink's formula and published by \citet{ekstrom12}, \citet{georgy13} and \citet{eggenberger21}.
	
	This paper is organised as it follows.
	We introduce at first a summary about our self-consistent m-CAK prescription, its foundations and range of validity, in Section~\ref{mcak}.
	We introduce later a brief recap about \textsc{Genec} in Section~\ref{genec}.
	The method for the implementation of our m-CAK prescription to derive our own recipe for the self-consistent mass-loss rate is outlined in Section~\ref{method}.
	We present our new self-consistent evolutionary tracks in Section~\ref{evolutionarytracks}, whereas the respective discussion is given in Section~\ref{structure}.
	Finally, the conclusions of our work are summarised in Section~\ref{conclusions}.
	
\section{Hydrodynamic m-CAK wind solutions}\label{mcak}
	\begin{figure*}[t!]
		\centering
		\includegraphics[width=0.45\linewidth]{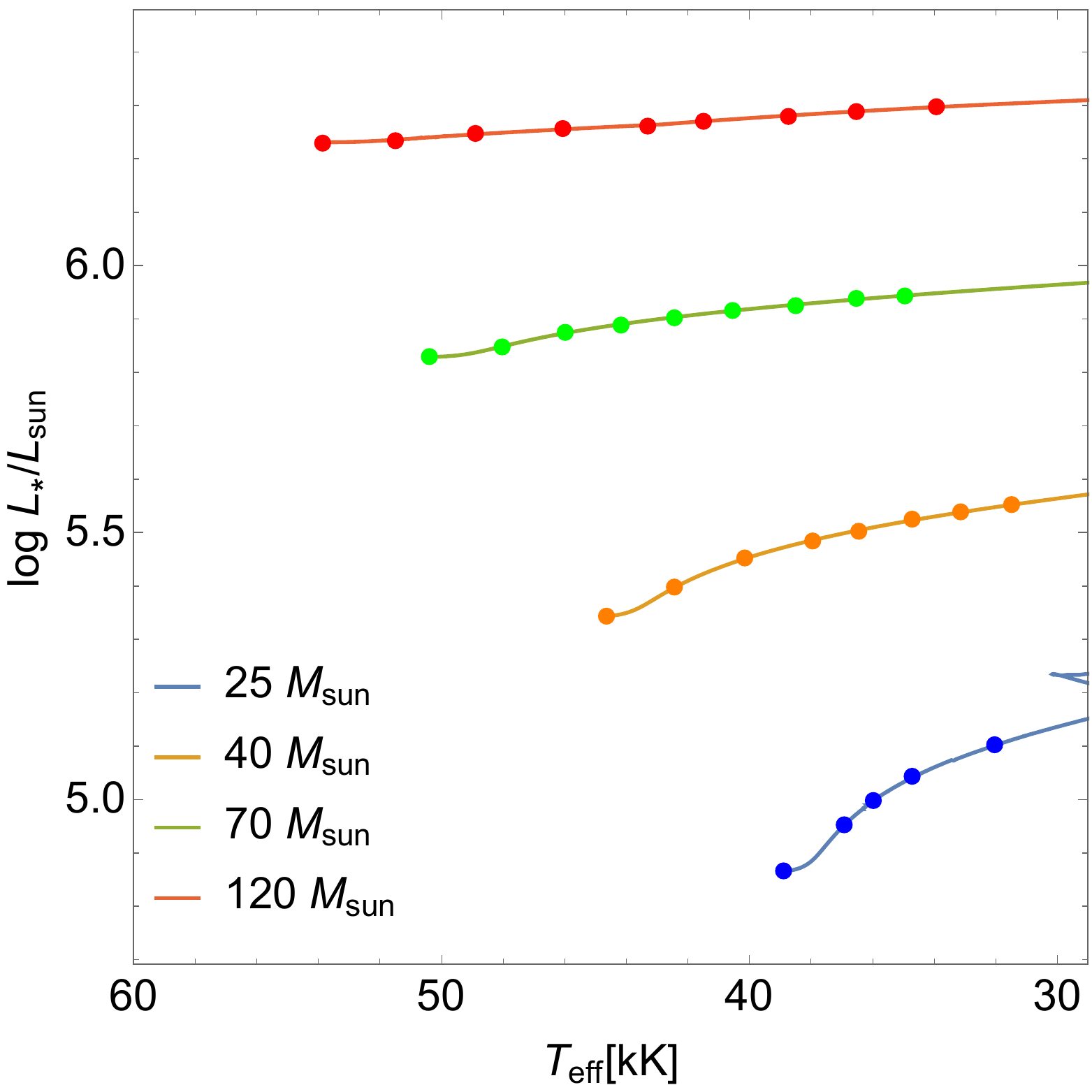}
		\hspace{1cm}
		\includegraphics[width=0.45\linewidth]{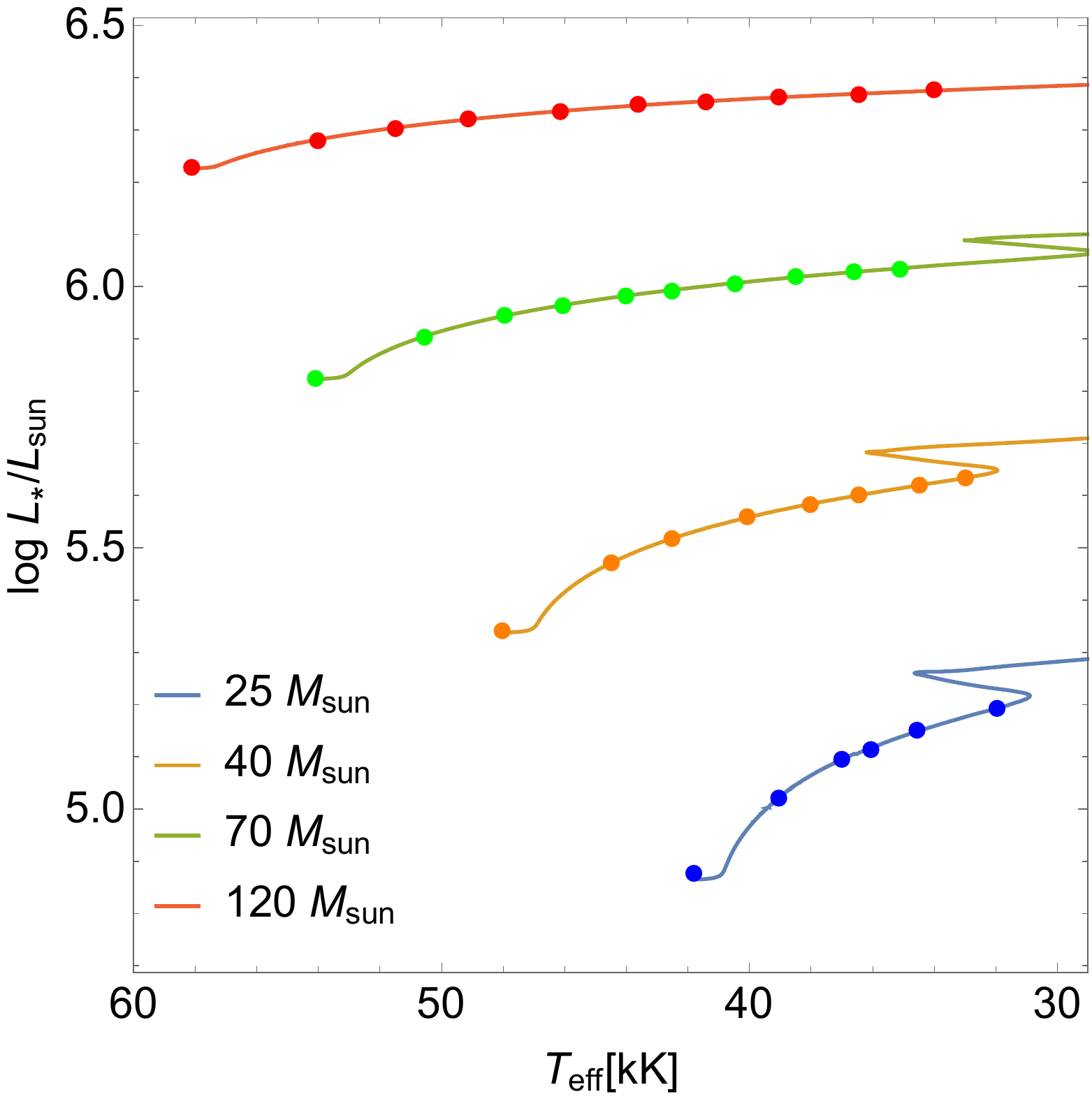}
		\caption{\small{Standard evolutionary tracks using the Vink's recipe for mass-loss rates, with $Z/Z_\odot=1.0$ (left panel) and $Z/Z_\odot=0.2$ (right panel). Location of the standard stars used to create the grid of self-consistent mass-loss rates are shown by the dots.}}
		\label{hrd_initial}
	\end{figure*}

	The initial study performed by \citet{alex19} provided an exhaustive analysis of the line-force parameters $k$, $\alpha$ and $\delta$ from the CAK theory \citep{cak,abbott82}
	\begin{equation}\label{mathcal}
		\mathcal M(t)=k\,t^{-\alpha}\left(N_{e,11}/W\right)^\delta\;,
	\end{equation}
	with $\mathcal M(t)$ being the force-multiplier (because it multiplies the acceleration due to electron scattering $g_\text{es}$ to generate the line-acceleration $g_\text{line}$), $t$ being the CAK optical depth and $N_{e,11}/W$ being the ionisation density.
	The value of these line-force parameters $(k,\alpha,\delta)$ are calculated from the wind hydrodynamics, which in turn are derived for the wind by solving the equation of motion using the code \textsc{HydWind} \citep{michel04}
	\begin{equation}\label{motion1}
		\varv\frac{d\varv}{dr}=-\frac{1}{\rho}\frac{dp}{dr}-\frac{GM_\text{eff}}{r^2}+g_\text{es}(r)\,k\,t^{-\alpha}\left(\frac{N_\text{e}}{W}\right)^\delta\;\;.
	\end{equation}
	
	We calculate the equation of motion adopting the correction by finite disk, following \citet{ppk}.
	This step includes some modifications on the original CAK theory from \citet[][thought for a source-point star]{cak}, and for that reason we currently talk about the modified-CAK (hereafter m-CAK) theory.
	
	Therefore, our calculation procedure of the line-force parameters is self-consistent with the hydrodynamics.
	As a consequence, new theoretical values for the wind parameters (mass-loss rate, terminal velocity) can be determined for any specific set of stellar parameters (effective temperature, surface gravity, stellar radius, abundances).
	In the particular case of mass-loss rates, self-consistent values have demonstrated to be in agreement with observably determined $\dot M$ when homogeneous wind --i.e, no clumping-- is assumed \citep[see Fig.~13 in][]{alex19}; whereas for the case of clumped winds, the clumping factor is found by set it as a free parameter in the spectral fitting.
	
	This spectral fitting for the self-consistent solutions is performed with the radiative transfer code FASTWIND \citep{santolaya97,puls05}, using as input a velocity profile generated from a formal solution of the equation of motion (Eq.~\ref{motion1}) instead of the classical $\beta$-law \citep{araya17}.
	Thus, it is possible to perform synthetic spectra with the wind parameters not longer free but dependent on the stellar ones, reducing then the number of free parameters.
	Based on this statement, we have implemented such self-consistent solutions to perform spectral fitting over a set of O-type type stars \citep{alex21b}, and obtaining then new stellar and wind parameters.

	Therefore, self-consistent m-CAK solutions provided by the procedures from \citet{alex19,alex21b} demonstrated to be reliable for O stars, despite of being based in a quasi-NLTE treatment for atomic populations.
	Studies aiming for a self-consistent prescription under a full NLTE treatment, such as the Lambert-procedure from \citet{alex21a}, have provided just slightly different results.
	For that reason, we state that theoretical values for mass-loss rate obtained from the self-consistent m-CAK solutions, $\dot M_\text{sc}$, are reliable to be implemented in studies about stellar evolution and to perform evolutionary tracks.
	
	As discussed in \citet{alex19} and \citet{alex21b}, validity of the m-CAK prescription is constrained to the range of temperatures where $(k,\alpha,\delta)$ can be treated as constant values.
	This region corresponds to stars with $T_\text{eff}\ge30$ kK and $\log g\ge3.2$, i.e., this region covers a relevant fraction of the lifetime of massive stars through their Main-Sequence stage.
	This is in agreement with the stated by \citet{puls08}, where it is established that the ``standard-model'' of line-driven theory is made for OB-stars excluding late evolutionary stages.
	Thus, m-CAK prescription cannot replace the theoretical mass-loss rates assumed for massive stars at stages such as LBVs or WRs, but still the change in the prescription for $\dot M$ \textit{at the beginning of the Main Sequence} will evidently have an impact over the future stellar conditions, as we expect to demonstrate in this work.

\section{Geneva evolution code}\label{genec}
	The Geneva evolution code (\textsc{Genec}) develops stellar evolutionary models to explain and predict the physical properties of massive stars.
	In such codes, the predicted effective temperature and luminosity (as a function of time) are used to build the tracks followed by the stars of different initial masses through the HR diagram.
	Other stellar properties, such as the mass and the abundances, are also calculated as a function of time.
	
	A detailed description of the code \textsc{Genec} can be found in \citet{eggenberger08}.
	For the present computations we considered only non-rotating stellar models.
	We used the same prescriptions as described in \citet{ekstrom12} –i.e., abundances from \citet{asplund05,asplund09}, opacities from \citet{iglesias96} and overshoot parameter~0.10– except for the mass-loss rates.
	The self-consistent m-CAK prescription has been implemented in \textsc{Genec}, for stars satisfying $T_\text{eff}\ge30$ kK and $\log g\ge3.2$.
	Below such thresholds, recipe for mass-loss rate returns to be Vink's formula.
	The metallicity dependence of this new mass-loss rate has been implemented too (see more details about this point in Section~\ref{method}).
	Except for this point, the models were computed with a moderate step overshoot applied at the Schwarzschild boundary for the convective cores.
	The initial compositions for the three metallicities have been taken the same as indicated in \citet{ekstrom12}, \citet{georgy13} and \citet{eggenberger21}.
	The models were computed without rotation. 

	The Geneva code in its non-rotating version gives similar results as would other codes if accounting for the same physical ingredients (as mass-loss, overshoot, and physical ingredients as the nuclear reaction rates, the equation of state, the opacity and the neutrino emissions).
	Comparisons between different codes and \textsc{Genec} can be found in \citet{meynet09}.
	
\section{Self-consistent mass-loss rates}\label{method}	
	As stated in \citet{alex19} and \citet{alex21b}, the self-consistency of our m-CAK prescription lies in the iterative process where both line-acceleration and wind hydrodynamics are simultaneously calculated for a set of stellar parameters.
	Because the running of these iterations for each one of the points compounding a \textsc{Genec} evolutionary track ($\sim5000$) is computationally unfeasible, we need to derive an easy mathematical formula.
	This formula needs to predict the value for $\dot M_\text{sc}$ using the stellar features of the star (temperature, radius, mass, metallicity and abundances), analogous to the Vink's formula \citep{vink01}.
	Nonetheless, for our case the formula does not need to assume any particular condition for the wind (such as a constant ratio for $\varv_\infty/\varv_\text{esc}$), because our wind parametrisation comes from our self-consistent prescription.
	
	For that reason, we extend the grid of self-consistent mass-loss rates, introduced in the Table~2 from \citet{alex21b}.
	Then we apply a statistical analysis to finally get a simple formula capable to fit $\dot M_\text{sc}$ in Section~\ref{statisticalfitting}.

\subsection{Extended grid of line-force solutions}
	We generate a grid of wind solutions wide enough to cover different stellar masses from their ZAMS stage until the end of the validity of the self-consistent prescription, assumed to be at $\log g\simeq3.2$ and $T_\text{eff}\simeq30$ kK as discussed in \citet{alex21b}.
	Therefore, to create that grid we have chosen some standard stars lying along the `classical' \citep[i.e., using the formulae from][for their mass-loss rates]{vink00,vink01} non-rotating evolutionary paths for initial masses of 25, 40, 70 and 120 $M_\odot$, and metallicities $Z/Z_\odot=1.0$, $Z/Z_\odot=0.5$ and $Z/Z_\odot=0.2$.
	Location of these standard stars across the Hertzsprung-Russell diagram (HRD), for the solar and the lowest metallicity, are shown in Fig.~\ref{hrd_initial}, whereas the names for the evolutionary tracks with their initial masses and metallicities are outlined in Table~\ref{trackseries}.

	\begin{table}[t!]
		\centering
		\caption{\small{Names of the classical evolutionary tracks for the sets of standard stars used to calculate $\dot M_\text{sc}$.}}
		\begin{tabular}{ccc}
			\hline\hline
			Name & Initial mass & Metallicity\\
			& $[M_\odot]$ & $[Z/Z_\odot]$\\
			\hline
			\texttt{P120z10} & 120 & 1.0\\
			\texttt{P070z10} & 70 & 1.0\\
			\texttt{P040z10} & 40 & 1.0\\
			\texttt{P025z10} & 25 & 1.0\\
			\texttt{P120z05} & 120 & 0.5\\
			\texttt{P070z05} & 70 & 0.5\\
			\texttt{P040z05} & 40 & 0.5\\
			\texttt{P025z05} & 25 & 0.5\\
			\texttt{P120z02} & 120 & 0.2\\
			\texttt{P070z02} & 70 & 0.2\\
			\texttt{P040z02} & 40 & 0.2\\
			\texttt{P025z02} & 25 & 0.2\\
			\hline
		\end{tabular}
		\label{trackseries}
	\end{table}
	\begin{table*}[t!]
		\centering
		\caption{\small{Self-consistent line-force parameters $(k,\alpha,\delta)$ for the set of stellar parameters of our standard stars from the classical evolutionary tracks (Table~\ref{trackseries}), together with their resulting terminal velocities and mass-loss rates ($\dot M_\text{sc}$). Ratios between self-consistent mass-loss rates with Vink's formula and KK17 values are also shown in the last columns.}}
		\resizebox{\textwidth}{!}{
		\begin{tabular}{cccccc|ccc|ccrr}
			\hline
			\hline
			Name & $T_\text{eff}$ & $\log g$ & $R_*$ & $M_*$ & $\log L_*$ & $k$ & $\alpha$ & $\delta$ & $\varv_\infty$ & $\log\dot M_\text{sc}$ & $\log\left(\frac{\dot M_\text{sc}}{\dot M_\text{Vink}}\right)$ & $\log\left(\frac{\dot M_\text{sc}}{\dot M_\text{KK17}}\right)$\\
			& $[\text{K}]$ & & $[R_\odot]$ & $[M_\odot]$ & $[L_\odot]$ & & & & [km s$^{-1}$] & [$M_\odot\,\text{yr}^{-1}$]\\
			\hline
			\texttt{P120z10-01} & 54\,000 & 4.15 & 15.0 & 120.0 & 6.23 & 0.121 & 0.571 & 0.026 & $3\,530\pm150$ & $-5.395\pm.072$ & $-0.339$ & $-0.080$\\
			\texttt{P120z10-02} & 51\,500 & 4.07 & 16.5 & 117.1 & 6.24 & 0.116 & 0.574 & 0.025 & $3\,420\pm150$ & $-5.411\pm.073$ & $-0.402$ & $-0.112$\\
			\texttt{P120z10-03} & 49\,000 & 3.95 & 18.5 & 111.8 & 6.25 & 0.102 & 0.584 & 0.022 & $3\,170\pm150$ & $-5.404\pm.067$ & $-0.461$ & $-0.121$\\
			\texttt{P120z10-04} & 46\,000 & 3.82 & 21.1 & 107.3 & 6.26 & 0.092 & 0.590 & 0.021 & $2\,930\pm130$ & $-5.415\pm.062$ & $-0.551$ & $-0.149$\\
			\texttt{P120z10-05} & 43\,500 & 3.70 & 24.0 & 105.4 & 6.27 & 0.084 & 0.604 & 0.027 & $2\,730\pm140$ & $-5.354\pm.068$ & $-0.530$ & $-0.104$\\
			\texttt{P120z10-06} & 41\,500 & 3.60 & 24.4 & 101.0 & 6.27 & 0.077 & 0.635 & 0.041 & $2\,560\pm130$ & $-5.304\pm.065$ & $-0.503$ & $-0.047$\\
			\texttt{P120z10-07} & 39\,000 & 3.45 & 30.7 & 97.4 & 6.28 & 0.069 & 0.670 & 0.054 & $2\,510\pm200$ & $-5.067\pm.068$ & $-0.278$ & $0.169$\\
			\texttt{P120z10-08} & 36\,500 & 3.33 & 34.9 & 95.1 & 6.29 & 0.061 & 0.681 & 0.063 & $2\,340\pm220$ & $-5.093\pm.100$ & $-0.303$ & $0.124$\\
			\texttt{P120z10-09} & 34\,000 & 3.20 & 40.8 & 93.2 & 6.30 & 0.054 & 0.689 & 0.083 & $2\,040\pm210$ & $-5.081\pm.113$ & $-0.243$ & $0.120$\\
			\hline
			\texttt{P070z10-01} & 50\,500 & 4.20 & 10.8 & 70.0 & 5.83 & 0.154 & 0.542 & 0.028 & $3\,110\pm130$ & $-5.946\pm.084$ & $-0.452$ & $0.021$\\
			\texttt{P070z10-02} & 48\,000 & 4.10 & 12.2 & 68.2 & 5.85 & 0.143 & 0.544 & 0.025 & $3\,000\pm130$ & $-5.962\pm.078$ & $-0.491$ & $-0.027$\\
			\texttt{P070z10-03} & 46\,000 & 4.00 & 13.5 & 66.2 & 5.87 & 0.129 & 0.547 & 0.021 & $2\,810\pm130$ & $-5.980\pm.076$ & $-0.553$ & $-0.077$\\
			\texttt{P070z10-04} & 44\,000 & 3.88 & 15.2 & 64.6 & 5.89 & 0.114 & 0.550 & 0.018 & $2\,570\pm130$ & $-5.977\pm.078$ & $-0.611$ & $-0.108$\\
			\texttt{P070z10-05} & 42\,500 & 3.80 & 16.6 & 63.5 & 5.90 & 0.106 & 0.554 & 0.019 & $2\,470\pm120$ & $-5.968\pm.083$ & $-0.631$ & $-0.115$\\
			\texttt{P070z10-06} & 40\,500 & 3.70 & 18.4 & 62.4 & 5.92 & 0.097 & 0.588 & 0.035 & $2\,420\pm130$ & $-5.801\pm.072$ & $-0.489$ & $0.019$\\
			\texttt{P070z10-07} & 38\,500 & 3.60 & 20.7 & 61.5 & 5.93 & 0.087 & 0.634 & 0.058 & $2\,490\pm150$ & $-5.591\pm.092$ & $-0.290$ & $0.213$\\
			\texttt{P070z10-08} & 36\,500 & 3.50 & 23.2 & 60.7 & 5.94 & 0.076 & 0.657 & 0.070 & $2\,450\pm170$ & $-5.557\pm.098$ & $-0.212$ & $0.231$\\
			\texttt{P070z10-09} & 35\,000 & 3.40 & 25.6 & 60.2 & 5.94 & 0.064 & 0.675 & 0.074 & $2\,380\pm230$ & $-5.551\pm.111$ & $-0.225$ & $0.237$\\
			\hline
			\texttt{P040z10-01} & 44\,500 & 4.25 & 7.8 & 40.0 & 5.34 & 0.241 & 0.479 & 0.024 & $2\,490\pm130$ & $-6.778\pm.118$ & $-0.487$ & $-0.022$\\
			\texttt{P040z10-02} & 42\,500 & 4.10 & 9.2 & 39.3 & 5.39 & 0.202 & 0.475 & 0.012 & $2\,280\pm130$ & $-6.819\pm.113$ & $-0.659$ & $-0.135$\\
			\texttt{P040z10-03} & 40\,000 & 3.95 & 11.0 & 38.4 & 5.45 & 0.176 & 0.487 & 0.013 & $2\,140\pm120$ & $-6.737\pm.113$ & $-0.677$ & $-0.150$\\
			\texttt{P040z10-04} & 38\,000 & 3.80 & 12.8 & 37.9 & 5.49 & 0.128 & 0.575 & 0.056 & $2\,230\pm120$ & $-6.274\pm.077$ & $-0.274$ & $0.247$\\
			\texttt{P040z10-05} & 36\,500 & 3.71 & 14.2 & 37.9 & 5.50 & 0.105 & 0.624 & 0.075 & $2\,370\pm130$ & $-6.086\pm.074$ & $-0.102$ & $0.418$\\
			\texttt{P040z10-06} & 34\,500 & 3.60 & 16.0 & 37.3 & 5.52 & 0.082 & 0.661 & 0.090 & $2\,400\pm140$ & $-6.048\pm.063$ & $0.026$ & $0.433$\\
			\texttt{P040z10-07} & 33\,000 & 3.50 & 19.4 & 37.1 & 5.54 & 0.067 & 0.677 & 0.096 & $2\,450\pm180$ & $-5.999\pm.053$ & $0.002$ & $0.442$\\
			\texttt{P040z10-08} & 31\,500 & 3.40 & 20.0 & 36.9 & 5.55 & 0.054 & 0.682 & 0.096 & $2\,200\pm280$ & $-6.158\pm.093$ & $-0.128$ & $0.266$\\
			\hline
			\texttt{P025z10-01} & 39\,000 & 4.28 & 6.0 & 25.0 & 4.87 & 0.362 & 0.442 & 0.027 & $2\,070\pm120$ & $-7.494\pm.121$ & $-0.386$ & $0.038$\\
			\texttt{P025z10-02} & 37\,000 & 4.10 & 7.3 & 24.8 & 4.95 & 0.223 & 0.486 & 0.036 & $2\,040\pm110$ & $-7.368\pm.068$ & $-0.424$ & $0.034$\\
			\texttt{P025z10-03} & 36\,000 & 4.00 & 8.1 & 24.7 & 5.00 & 0.178 & 0.515 & 0.047 & $2\,010\pm110$ & $-7.224\pm.063$ & $-0.357$ & $0.096$\\
			\texttt{P025z10-04} & 34\,500 & 3.90 & 9.0 & 24.5 & 5.04 & 0.122 & 0.613 & 0.094 & $2\,270\pm120$ & $-6.769\pm.106$ & $0.039$ & $0.486$\\
			\texttt{P025z10-05} & 32\,000 & 3.70 & 11.5 & 24.3 & 5.10 & 0.070 & 0.663 & 0.106 & $2\,320\pm120$ & $-6.741\pm.111$ & $0.024$ & $0.416$\\
			\hline
		\end{tabular}}
		\label{standardtable}
	\end{table*}
		
	Given the solutions for line-force parameters employed by \citet{alex19,alex21b}, we determined new self-consistent values for mass-loss rate for these selected standard stars.
	These standard stars are named as \texttt{-01}, \texttt{-02}, \texttt{-03}, etc., as a function of the respective standard track from Table~\ref{trackseries}, and they are selected aiming to keep an almost constant distance in the space of $T_\text{eff}$ and $\log g$.
	The total number of standard stars per each track varies, depending on how long is the range of temperatures and gravities before reaching the thresholds for the m-CAK prescription.
	 Results are shown in Table~\ref{standardtable} for solar-metallicity ($Z/Z_\odot=1.0$), in Table~\ref{standardtablez07} for $Z/Z_\odot=0.5$ and in Table~\ref{standardtablez03} for $Z/Z_\odot=0.2$.
	Error bars associated to the wind parameters, $\varv_\infty$ and $\dot M_\text{sc}$, are based on the uncertainties of the line-force parameters as described in \citet{alex21b}.
	Comparison between these new self-consistent values for $\dot M$ and those used previously, determined by the Vink's formula, are also included.
	Besides, we add an extra column comparing our $\dot M_\text{sc}$ with those theoretical values provided by \citet[][hereafter KK17, see their Eq.~11]{kk17} and \citet[][hereafter KK18, see their Eq.~1]{kk18} because their formulae is also based on a theoretical prescription for the stellar wind, although both formulae describe mass-loss rate only as a function of stellar luminosity.
	
	\begin{table*}[t!]
		\centering
		\caption{\small{Analogous to Table~\ref{standardtable}, grid of standard stars extracted from the classical evolutionary tracks for $Z/Z_\odot=0.5$.}}
		\resizebox{\textwidth}{!}{
		\begin{tabular}{cccccc|ccc|ccrr}
			\hline
			\hline
			Name & $T_\text{eff}$ & $\log g$ & $R_*$ & $M_*$ & $\log L_*$ & $k$ & $\alpha$ & $\delta$ & $\varv_\infty$ & $\log\dot M_\text{sc}$ & $\log\left(\frac{\dot M_\text{sc}}{\dot M_\text{Vink}}\right)$ & $\log\left(\frac{\dot M_\text{sc}}{\dot M_\text{KK18}}\right)$\\
			& $[\text{K}]$ & & $[R_\odot]$ & $[M_\odot]$ & $[L_\odot]$ & & & & [km s$^{-1}$] & [$M_\odot\,\text{yr}^{-1}$]\\
			\hline
			\texttt{P120z05-01} & 56\,000 & 4.24 & 13.8 & 120.0 & 6.23 & 0.139 & 0.514 & 0.028 & $3\,110\pm120$ & $-5.620\pm.065$ & $-0.260$ & $-0.148$\\
			\texttt{P120z05-02} & 53\,000 & 4.11 & 15.8 & 117.4 & 6.25 & 0.119 & 0.537 & 0.031 & $2\,990\pm130$ & $-5.536\pm.065$ & $-0.288$ & $-0.097$\\
			\texttt{P120z05-03} & 51\,000 & 4.00 & 17.4 & 111.2 & 6.27 & 0.102 & 0.553 & 0.028 & $2\,820\pm130$ & $-5.500\pm.063$ & $-0.328$ & $-0.094$\\
			\texttt{P120z05-04} & 47\,500 & 3.86 & 20.6 & 110.5 & 6.29 & 0.091 & 0.559 & 0.023 & $2\,640\pm120$ & $-5.510\pm.057$ & $-0.434$ & $-0.137$\\
			\texttt{P120z05-05} & 44\,000 & 3.70 & 24.4 & 108.9 & 6.30 & 0.082 & 0.557 & 0.020 & $2\,300\pm120$ & $-5.532\pm.059$ & $-0.513$ & $-0.175$\\
			\texttt{P120z05-06} & 42\,000 & 3.61 & 27.0 & 108.4 & 6.31 & 0.078 & 0.569 & 0.026 & $2\,200\pm130$ & $-5.474\pm.060$ & $-0.472$ & $-0.134$\\
			\texttt{P120z05-07} & 40\,000 & 3.51 & 30.0 & 106.3 & 6.32 & 0.071 & 0.617 & 0.049 & $2\,240\pm150$ & $-5.263\pm.050$ & $-0.266$ & $0.061$\\
			\texttt{P120z05-08} & 37\,000 & 3.36 & 35.5 & 105.4 & 6.33 & 0.061 & 0.658 & 0.068 & $2\,190\pm200$ & $-5.137\pm.051$ & $-0.121$ & $0.171$\\
			\texttt{P120z05-09} & 34\,000 & 3.20 & 42.4 & 104.0 & 6.33 & 0.050 & 0.681 & 0.082 & $2\,020\pm190$ & $-5.137\pm.080$ & $-0.068$ & $0.171$\\
			\hline
			\texttt{P070z05-01} & 52\,000 & 4.27 & 10.1 & 70.0 & 5.82 & 0.180 & 0.495 & 0.036 & $2\,740\pm120$ & $-6.111\pm.084$ & $-0.240$ & $0.036$\\
			\texttt{P070z05-02} & 50\,000 & 4.17 & 11.3 & 68.9 & 5.85 & 0.156 & 0.510 & 0.034 & $2\,690\pm130$ & $-6.063\pm.087$ & $-0.287$ & $0.034$\\
			\texttt{P070z05-03} & 48\,000 & 4.05 & 12.8 & 67.1 & 5.89 & 0.132 & 0.521 & 0.030 & $2\,540\pm120$ & $-6.034\pm.073$ & $-0.370$ & $-0.002$\\
			\texttt{P070z05-04} & 46\,000 & 3.95 & 14.3 & 66.5 & 5.92 & 0.121 & 0.519 & 0.024 & $2\,360\pm120$ & $-6.066\pm.082$ & $-0.474$ & $-0.084$\\
			\texttt{P070z05-05} & 44\,000 & 3.85 & 16.0 & 66.1 & 5.93 & 0.115 & 0.507 & 0.017 & $2\,150\pm120$ & $-6.127\pm.087$ & $-0.587$ & $-0.161$\\
			\texttt{P070z05-06} & 42\,000 & 3.75 & 17.9 & 65.8 & 5.95 & 0.110 & 0.506 & 0.012 & $2\,020\pm120$ & $-6.112\pm.094$ & $-0.607$ & $-0.179$\\
			\texttt{P070z05-07} & 39\,500 & 3.62 & 20.6 & 64.6 & 5.96 & 0.094 & 0.565 & 0.043 & $2\,020\pm130$ & $-5.817\pm.077$ & $-0.332$ & $0.099$\\
			\texttt{P070z05-08} & 37\,000 & 3.49 & 23.8 & 63.9 & 5.98 & 0.076 & 0.632 & 0.074 & $2\,130\pm160$ & $-5.577\pm.064$ & $-0.088$ & $0.306$\\
			\texttt{P070z05-09} & 35\,000 & 3.38 & 26.9 & 63.4 & 5.99 & 0.063 & 0.662 & 0.086 & $2\,120\pm200$ & $-5.542\pm.053$ & $-0.030$ & $0.325$\\
			\hline
			\texttt{P040z05-01} & 46\,000 & 4.31 & 7.3 & 40.0 & 5.34 & 0.281 & 0.453 & 0.039 & $2\,270\pm120$ & $-6.837\pm.122$ & $-0.270$ & $0.100$\\
			\texttt{P040z05-02} & 44\,000 & 4.16 & 8.7 & 39.9 & 5.41 & 0.234 & 0.451 & 0.027 & $2\,110\pm120$ & $-6.850\pm.105$ & $-0.443$ & $-0.028$\\
			\texttt{P040z05-03} & 42\,000 & 4.01 & 10.2 & 38.9 & 5.47 & 0.192 & 0.452 & 0.014 & $1\,960\pm120$ & $-6.868\pm.106$ & $-0.593$ & $-0.145$\\
			\texttt{P040z05-04} & 40\,000 & 3.88 & 11.8 & 38.5 & 5.51 & 0.180 & 0.451 & 0.014 & $1\,760\pm120$ & $-6.814\pm.109$ & $-6.199$ & $-0.157$\\
			\texttt{P040z05-05} & 38\,000 & 3.76 & 13.5 & 38.3 & 5.53 & 0.144 & 0.494 & 0.036 & $1\,740\pm120$ & $-6.611\pm.096$ & $-0.452$ & $0.013$\\
			\texttt{P040z05-06} & 36\,000 & 3.64 & 15.5 & 38.2 & 5.56 & 0.098 & 0.614 & 0.086 & $2\,070\pm130$ & $-6.082\pm.100$ & $0.062$ & $0.493$\\
			\texttt{P040z05-07} & 34\,000 & 3.52 & 17.7 & 37.9 & 5.58 & 0.073 & 0.660 & 0.103 & $2\,130\pm150$ & $-6.016\pm.087$ & $0.138$ & $0.526$\\
			\texttt{P040z05-08} & 32\,000 & 3.40 & 20.3 & 37.8 & 5.59 & 0.056 & 0.674 & 0.106 & $2\,050\pm240$ & $-6.113\pm.068$ & $0.073$ & $0.412$\\
			\hline
			\texttt{P025z05-01} & 40\,000 & 4.33 & 5.7 & 25.0 & 4.86 & 0.375 & 0.439 & 0.041 & $2\,030\pm110$ & $-7.501\pm.111$ & $-0.145$ & $0.226$\\
			\texttt{P025z05-02} & 38\,000 & 4.13 & 7.1 & 24.8 & 4.98 & 0.180 & 0.499 & 0.043 & $2\,120\pm100$ & $-7.426\pm.057$ & $-0.291$ & $0.104$\\
			\texttt{P025z05-03} & 36\,000 & 3.95 & 8.7 & 24.6& 5.06 & 0.083 & 0.568 & 0.051 & $2\,300\pm100$ & $-7.360\pm.045$ & $-0.371$ & $-0.038$\\
			\texttt{P025z05-04} & 34\,000 & 3.80 & 10.3 & 24.3 & 5.11 & 0.023 & 0.659 & 0.052 & $2\,960\pm130$ & $-7.565\pm.104$ & $-0.631$ & $-0.249$\\
			\texttt{P025z05-05} & 32\,000 & 3.66 & 12.2 & 24.2 & 5.15 & 0.013 & 0.677 & 0.025 & $3\,280\pm200$ & $-7.769\pm.137$ & $-0.848$ & $-0.519$\\
			\hline
		\end{tabular}}
		\label{standardtablez07}
	\end{table*}
	\begin{table*}[t!]
		\centering
		\caption{\small{Analogous to Tables~\ref{standardtable} and \ref{standardtablez07}, grid of standard stars extracted from the classical evolutionary tracks for $Z/Z_\odot=0.2$.}}
		\resizebox{\textwidth}{!}{
		\begin{tabular}{cccccc|ccc|ccrr}
			\hline
			\hline
			Name & $T_\text{eff}$ & $\log g$ & $R_*$ & $M_*$ & $\log L_*$ & $k$ & $\alpha$ & $\delta$ & $\varv_\infty$ & $\log\dot M_\text{sc}$ & $\log\left(\frac{\dot M_\text{sc}}{\dot M_\text{Vink}}\right)$ & $\log\left(\frac{\dot M_\text{sc}}{\dot M_\text{KK18}}\right)$\\
			& $[\text{K}]$ & & $[R_\odot]$ & $[M_\odot]$ & $[L_\odot]$ & & & & [km s$^{-1}$] & [$M_\odot\,\text{yr}^{-1}$]\\
			\hline
			\texttt{P120z02-01} & 58\,000 & 4.30 & 12.8 & 120.0 & 6.23 & 0.199 & 0.383 & 0.012 & $2\,230\pm120$ & $-6.155\pm.136$ & $-0.543$ & $-0.495$\\
			\texttt{P120z02-02} & 54\,000 & 4.10 & 15.8 & 117.3 & 6.28 & 0.120 & 0.481 & 0.035 & $2\,310\pm130$ & $-5.718\pm.080$ & $-0.176$ & $-0.143$\\
			\texttt{P120z02-03} & 51\,500 & 4.00 & 17.8 & 116.3 & 6.30 & 0.104 & 0.504 & 0.037 & $2\,300\pm130$ & $-5.652\pm.088$ & $-0.178$ & $-0.111$\\
			\texttt{P120z02-04} & 49\,000 & 3.90 & 20.0 & 115.4 & 6.32 & 0.093 & 0.515 & 0.035 & $2\,230\pm120$ & $-5.641\pm.081$ & $-0.204$ & $-0.134$\\
			\texttt{P120z02-05} & 46\,000 & 3.77 & 23.1 & 114.7 & 6.34 & 0.083 & 0.513 & 0.026 & $2\,030\pm120$ & $-5.698\pm.090$ & $-0.273$ & $-0.224$\\
			\texttt{P120z02-06} & 43\,500 & 3.65 & 26.2 & 114.1 & 6.35 & 0.078 & 0.498 & 0.015 & $1\,790\pm120$ & $-5.778\pm.083$ & $-0.404$ & $-0.321$\\
			\texttt{P120z02-07} & 41\,500 & 3.56 & 29.3 & 113.6 & 6.36 & 0.075 & 0.502 & 0.014 & $1\,710\pm130$ & $-5.741\pm.094$ & $-0.423$ & $-0.301$\\
			\texttt{P120z02-08} & 39\,000 & 3.45 & 33.2 & 113.1 & 6.36 & 0.070 & 0.570 & 0.054 & $1\,770\pm160$ & $-5.399\pm.075$ & $-0.273$ & $0.041$\\
			\texttt{P120z02-09} & 36\,500 & 3.32 & 38.4 & 112.6 & 6.37 & 0.057 & 0.640 & 0.081 & $1\,890\pm200$ & $-5.173\pm.070$ & $-0.193$ & $0.250$\\
			\texttt{P120z02-10} & 34\,000 & 3.19 & 44.5 & 112.2 & 6.38 & 0.046 & 0.670 & 0.095 & $1\,690\pm180$ & $-5.291\pm.096$ & $0.089$ & $0.115$\\
			\hline
			\texttt{P070z02-01} & 54\,000 & 4.35 & 9.3 & 70.0 & 5.82 & 0.354 & 0.323 & 0.015 & $1\,810\pm110$ & $-5.830\pm.158$ & $0.432$ & $0.524$\\
			\texttt{P070z02-02} & 50\,500 & 4.14 & 11.7 & 69.0 & 5.90 & 0.166 & 0.457 & 0.044 & $2\,100\pm110$ & $-6.112\pm.096$ & $-0.200$ & $0.107$\\
			\texttt{P070z02-03} & 48\,000 & 4.00 & 13.6 & 68.3 & 5.94 & 0.131 & 0.482 & 0.041 & $2\,030\pm110$ & $-6.134\pm.087$ & $-0.245$ & $0.017$\\
			\texttt{P070z02-04} & 46\,000 & 3.91 & 15.1 & 67.9 & 5.96 & 0.121 & 0.479 & 0.034 & $1\,920\pm110$ & $-6.186\pm.082$ & $-0.347$ & $-0.069$\\
			\texttt{P070z02-05} & 44\,000 & 3.81 & 16.9 & 67.6 & 5.98 & 0.117 & 0.464 & 0.022 & $1\,750\pm110$ & $-6.256\pm.093$ & $-0.468$ & $-0.173$\\
			\texttt{P070z02-06} & 42\,500 & 3.74 & 18.3 & 67.4 & 5.99 & 0.109 & 0.462 & 0.014 & $1\,690\pm120$ & $-6.310\pm.093$ & $-0.666$ & $-0.244$\\
			\texttt{P070z02-07} & 40\,500 & 3.64 & 20.5 & 67.1 & 6.01 & 0.104 & 0.460 & 0.014 & $1\,550\pm120$ & $-6.308\pm.088$ & $-0.568$ & $-0.275$\\
			\texttt{P070z02-08} & 38\,500 & 3.54 & 23.0 & 66.9 & 6.02 & 0.095 & 0.491 & 0.036 & $1\,500\pm130$ & $-6.129\pm.080$ & $-0.388$ & $-0.113$\\
			\texttt{P070z02-09} & 36\,500 & 3.44 & 25.8 & 66.7 & 6.03 & 0.073 & 0.601 & 0.085 & $1\,750\pm160$ & $-5.662\pm.104$ & $-0.109$ & $0.337$\\
			\texttt{P070z02-10} & 35\,000 & 3.36 & 28.2 & 66.6 & 6.03 & 0.061 & 0.645 & 0.101 & $1\,840\pm180$ & $-5.562\pm.116$ & $0.180$ & $0.437$\\
			\hline
			\texttt{P040z02-01} & 48\,000 & 4.40 & 6.8 & 40.0 & 5.34 & 0.501 & 0.347 & 0.051 & $1\,700\pm120$ & $-7.192\pm.220$ & $-0.280$ & $-0.025$\\
			\texttt{P040z02-02} & 44\,500 & 4.11 & 9.2 & 39.6 & 5.47 & 0.235 & 0.427 & 0.043 & $1\,770\pm120$ & $-6.845\pm.158$ & $-0.245$ & $0.102$\\
			\texttt{P040z02-03} & 42\,500 & 3.98 & 10.6 & 39.3 & 5.52 & 0.209 & 0.424 & 0.029 & $1\,650\pm120$ & $-6.861\pm.122$ & $-0.357$ & $0.002$\\
			\texttt{P040z02-04} & 40\,000 & 3.84 & 12.5 & 39.2 & 5.56 & 0.160 & 0.444 & 0.027 & $1\,600\pm110$ & $-6.850\pm.099$ & $-0.418$ & $-0.055$\\
			\texttt{P040z02-05} & 38\,000 & 3.72 & 14.3 & 39.1 & 5.58 & 0.108 & 0.486 & 0.035 & $1\,640\pm110$ & $-6.824\pm.065$ & $-0.423$ & $-0.063$\\
			\texttt{P040z02-06} & 36\,500 & 3.63 & 15.8 & 39.0 & 5.60 & 0.075 & 0.530 & 0.044 & $1\,710\pm110$ & $-6.796\pm.084$ & $-0.398$ & $-0.069$\\
			\texttt{P040z02-07} & 34\,500 & 3.51 & 18.2 & 38.9 & 5.62 & 0.076 & 0.613 & 0.107 & $1\,740\pm140$ & $-6.159\pm.106$ & $0.239$ & $0.534$\\
			\texttt{P040z02-08} & 33\,000 & 3.42 & 20.1 & 38.9 & 5.63 & 0.060 & 0.654 & 0.119 & $1\,810\pm170$ & $-6.088\pm.114$ & $0.315$ & $0.588$\\
			\hline
			\texttt{P025z02-01} & 42\,000 & 4.40 & 5.2 & 25.0 & 4.88 & 0.638 & 0.275 & 0.014 & $1\,360\pm120$ & $-8.962\pm.333$ & $-1.324$ & $-1.015$\\
			\texttt{P025z02-02} & 39\,000 & 4.13 & 7.1 & 24.9 & 5.02 & 0.140 & 0.526 & 0.068 & $2\,140\pm110$ & $-7.323\pm.076$ & $0.042$ & $0.386$\\
			\texttt{P025z02-03} & 37\,000 & 3.96 & 8.6 & 24.9 & 5.09 & 0.054 & 0.610 & 0.074 & $2\,490\pm120$ & $-7.345\pm.067$ & $-0.119$ & $0.246$\\
			\texttt{P025z02-04} & 36\,000 & 3.90 & 9.3 & 24.8 & 5.12 & 0.035 & 0.643 & 0.074 & $2\,730\pm120$ & $-7.393\pm.079$ & $-0.192$ & $0.147$\\
			\texttt{P025z02-05} & 34\,500 & 3.79 & 10.5 & 24.8 & 5.15 & 0.020 & 0.669 & 0.067 & $2\,870\pm180$ & $-7.556\pm.094$ & $-0.377$ & $-0.067$\\
			\texttt{P025z02-06} & 32\,000 & 3.61 & 12.9 & 24.7 & 5.19 & 0.015 & 0.659 & 0.029 & $2\,790\pm230$ & $-7.676\pm.123$ & $-0.512$ & $-0.254$\\
			\hline
		\end{tabular}}
		\label{standardtablez03}
	\end{table*}

\subsection{Statistical fit}\label{statisticalfitting}
	Given the grid of values for theoretical self-consistent mass-loss rate $\dot M_\text{sc}$, outlined in Tables~\ref{standardtable}, \ref{standardtablez07} and \ref{standardtablez03}, we proceed with the statistical analysis to generate an easy-to-implement formula, comparable to the known Vink's formula.
	For that reason, we explore three possibilities for fitting:
	\begin{itemize}
		\item\textit{Linear fitting}
		\begin{equation}
			Y(x_1,x_2,...,x_N)=A_0+\sum_{i=1}^N A_ix_i\;,
		\end{equation}
		\item\textit{Quadratic fitting}
		\begin{equation}
			Y(x_1,x_2,...,x_N)=A_0+\sum_{i=1}^N A_ix_i+\sum_{i=1}^N B_ix_i^2\;,
		\end{equation}
		\item\textit{Intervariable fitting}
		\begin{equation}
			Y(x_1,x_2,...,x_N)=A_0+\sum_{i=1}^N A_ix_i+\sum_{i,j=1}^N B_{ij}x_ix_j\;,
		\end{equation}
	\end{itemize}
	where $Y$ is our fitted mass-loss rate (expressed either as $\dot M$ or $\log\dot M$), and $x_1,...,x_N$ are our individual stellar parameters (temperature, gravity, radius and metallicity).
	Linear fit was previously implemented by KK17 and \citet{alex19}, whereas quadratic fitting is used by Vink's formula.
	Besides these options, we include the intervariable fitting in order to determine any potential inter-dependence among the individual stellar parameters.
	
	Per each one of the three introduced alternatives, stellar parameters are evaluated in their linear ($x_i$), inverse ($1/x_i$) and logarithmical ($\log x_i$) scales, looking for the combination which better satisfies our criteria.
	Criteria to select a good fit are:
	\begin{itemize}
		\item A closest value to 1 for the coefficient of determination $R$-squared ($R^2$).
		\item Higher values on $t$-Statistic for each linear parameter.
		\item A lower dispersion between the real (i.e., the value from the Tables~\ref{standardtable}, \ref{standardtablez07} and \ref{standardtablez03}) and predicted $\dot M_\text{sc}$, weighted by the error bars $\Delta\dot M$.
		Then we define our \textit{weighted deviation} as
		\begin{equation}
			\Sigma_\text{w}=\left|\frac{\dot M_\text{sc,predicted}-\dot M_\text{sc,real}}{\Delta\dot M}\right|\;,
		\end{equation}
	\end{itemize}
	
	The advantage for adopting this weighted deviation is, we can gauge the difference between predicted and real mass-loss rates in terms relative to the intrinsic uncertainty for our results as outlined in \citet{alex21b}.
	Given the definition of $\Sigma_\text{w}$, it becomes evident that a value lower than 1 is the most idealistic result because it implies the predicted response of our fit lies below the intrinsic error of our models.
	This means, the more reduced spreading of the deviations the better the fit.
	Therefore, we expect to find a dispersion with the highest percentage of the sample below the thresholds 2 and 3.
	
	\begin{table}[t!]
		\centering
		\caption{\small{Coefficient of determination $R^2$ obtained for our three alternatives for fitting.}}
		\begin{tabular}{ccc}
			\hline\hline
			Fitting & $R^2$ & $\log(1-R^2)$\\
			\hline
			Linear & 0.99872 & $-2.8942$\\
			Quadratic & 0.99887 & $-2.9466$\\
			Intervariable & 0.99895 & $-2.9777$\\
			\hline
		\end{tabular}
		\label{r2table}
	\end{table}
	
	\begin{figure}[t!]
		\centering
		\includegraphics[width=0.95\linewidth]{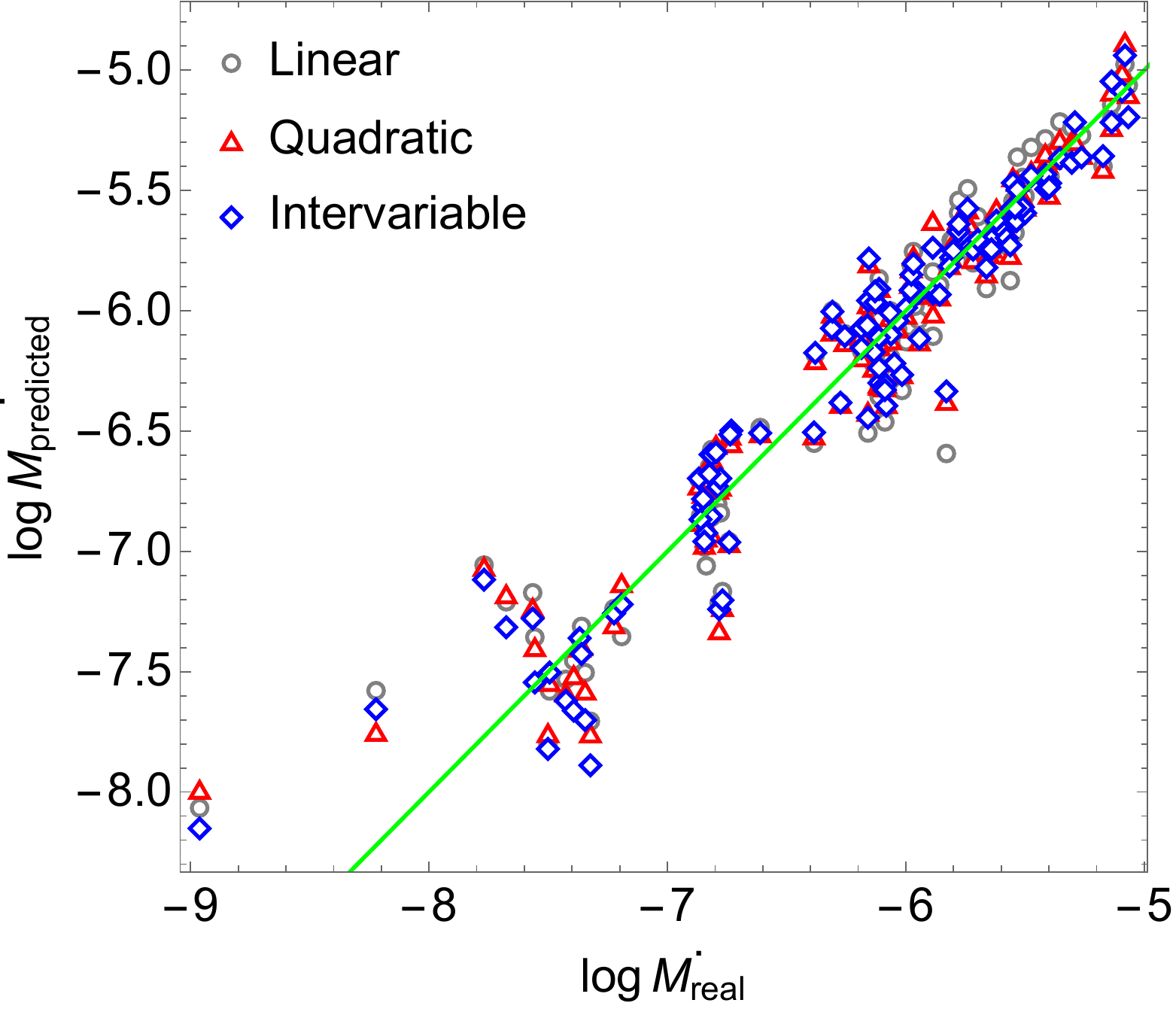}
		\caption{\small{Distribution of the obtained mass-loss rate values from our three alternatives for fitting (linear, quadratic and intervariable) in comparison with the true theoretical mass-loss rate coming from self-consistent wind solutions.}}
		\label{mdot_fitting}
	\end{figure}
	
	\begin{figure*}[t!]
		\centering
		\includegraphics[width=0.32\linewidth]{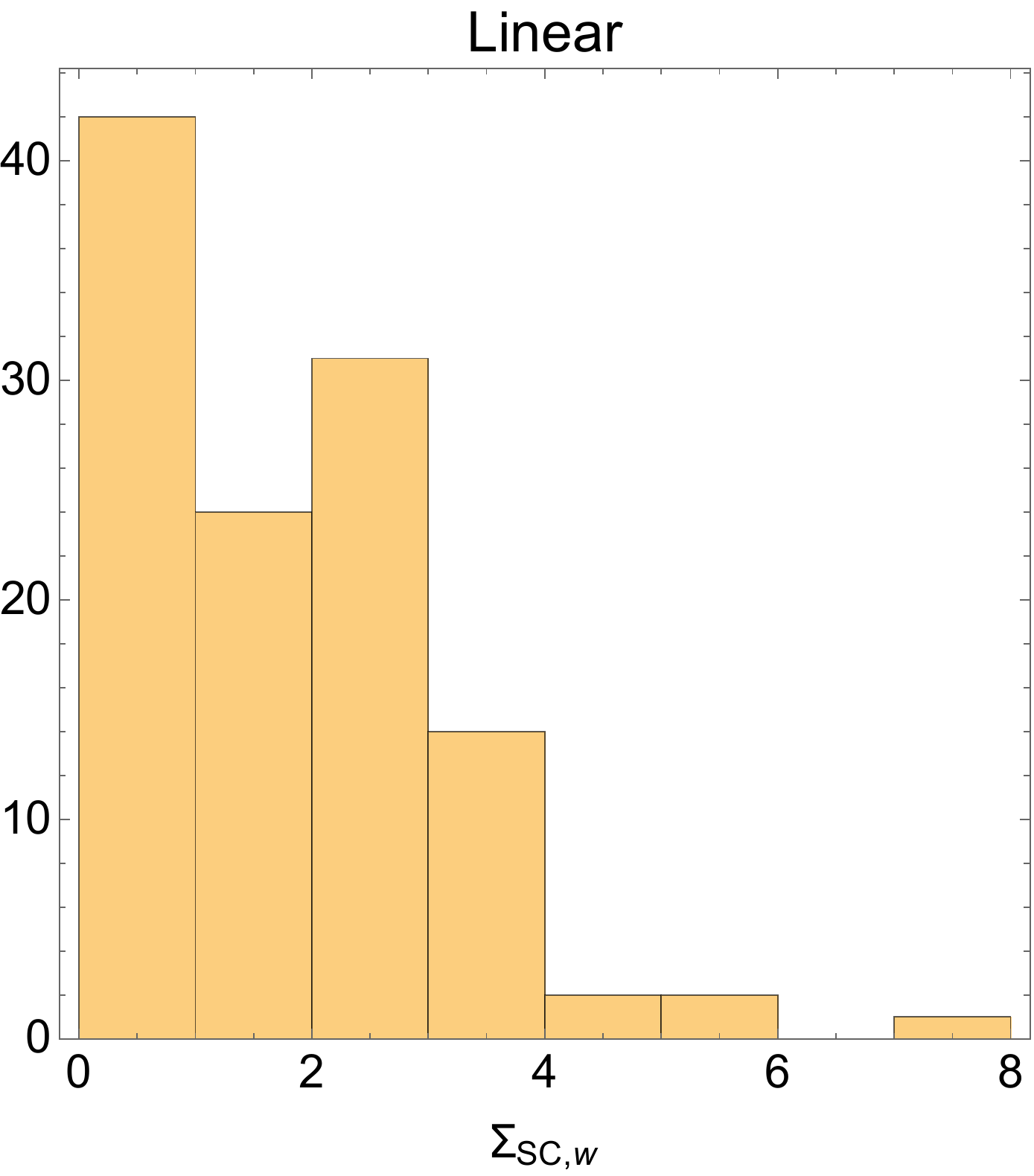}
		\hspace{2mm}
		\includegraphics[width=0.32\linewidth]{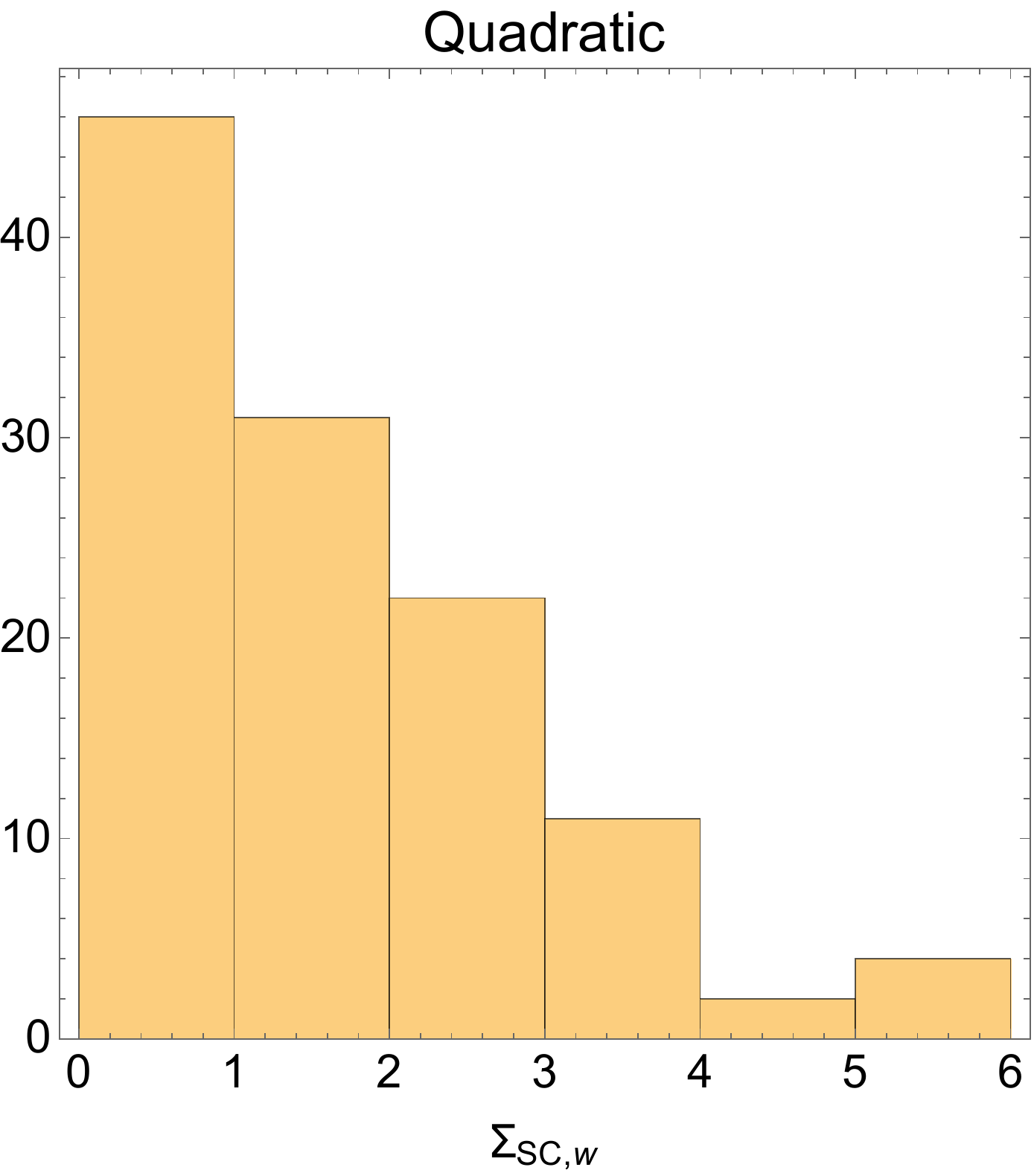}
		\hspace{2mm}
		\includegraphics[width=0.32\linewidth]{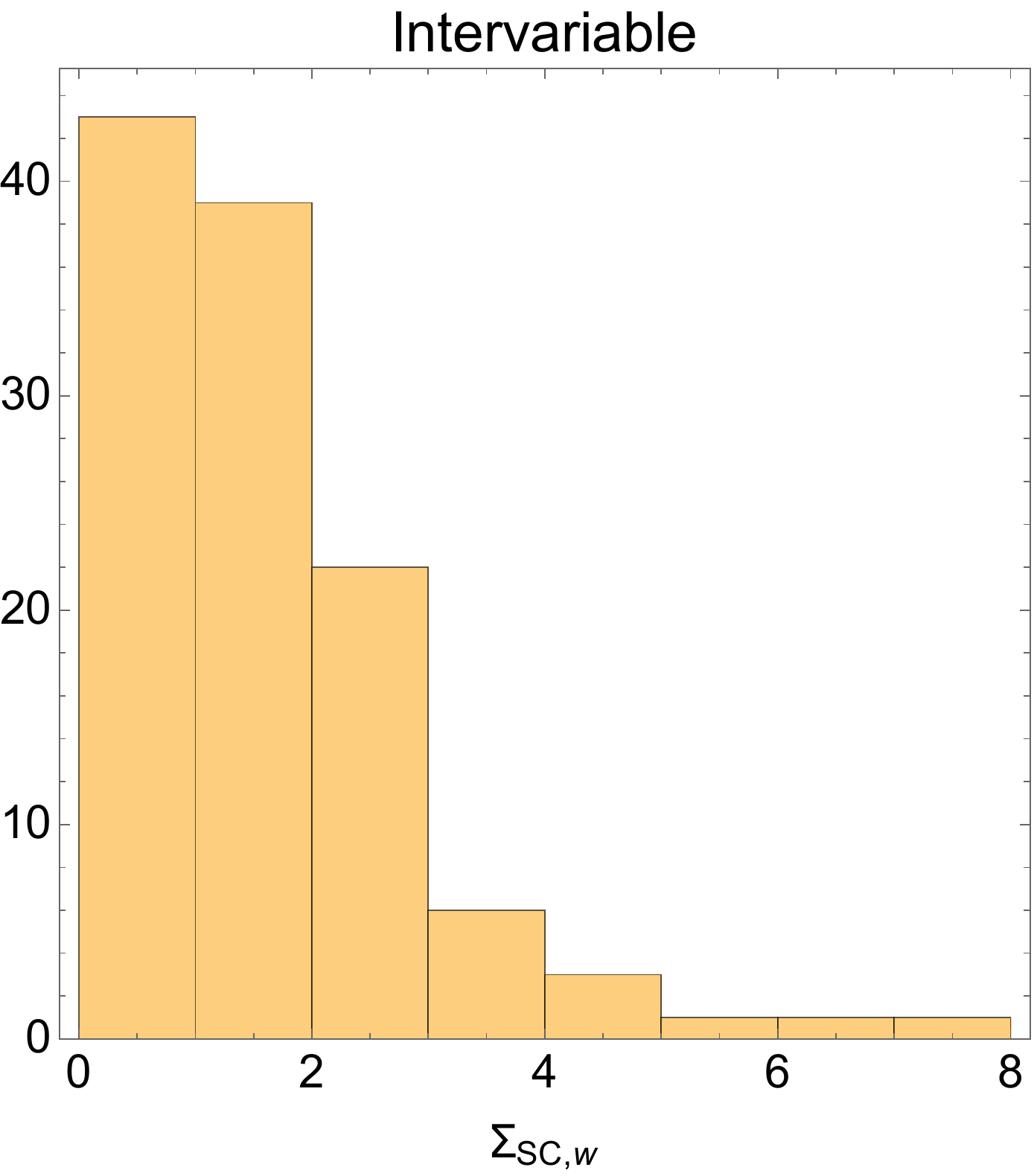}
		\caption{\small{Histograms for our three attempts of fitting: linear, quadratic and intervariable.}}
		\label{histograms}
	\end{figure*}

	Combining the total of stellar wind models from Tables~\ref{standardtable}, \ref{standardtablez07} and \ref{standardtablez03}, with those coming from Tables~1 and 2 from \citet{alex21b}, we get a total set of 116 self-consistent mass-loss rates.
	The result of our three attempts for fitting are presented in Table~\ref{r2table}.
	It is observed that our three attempts presents a good prediction potential based on their $R^2\simeq1$, even though the intervariable fitting presents the closest approximation to the unity.
	This can be also appreciated in Fig.~\ref{mdot_fitting}, where the three attempts closely match the ideal $\dot M_\text{predicted}=\dot M_\text{real}$, with the intervariable fitting slightly standing out between the other two.
	If we consider the normalisation by the error bars of the self-consistent mass-loss rates, and we plot histograms for the distribution of the $\Sigma_\text{w}$ factor (see Fig.~\ref{histograms}), we observe a smaller spreading for the quadratic fitting but the intervariable fitting keeps the highest percentage of the sample below $\Sigma_\text{w}\le3$, with a $\sim90\%$.
	This is illustrated in Table~\ref{sigmathreshold}, where both the quadratic and the intervariable fittings show a big percentage of the sample satisfying the thresholds $\Sigma_\text{w}\le2$ and $\Sigma_\text{w}\le3$ (over $68\%$ and $87\%$ respectively), with a slight advantage for the intervariable fitting.
	Because this result is in agreement with the statistics found for $R^2$, we decide to employ the intervariable fitting.
	
	\begin{table}[t!]
		\centering
		\caption{\small{Coefficient of determination $R^2$ obtained for our three alternatives for fitting.}}
		\begin{tabular}{cccc}
			\hline\hline
			Fitting & $\%[\Sigma_\text{w}\le1]$ & $\%[\Sigma_\text{w}\le2]$ & $\%[\Sigma_\text{w}\le3]$\\
			\hline
			Linear & $34.5$ & $55.2$ & $81.9$\\
			Quadratic & $39.7$ & $68.1$ & $87.1$\\
			Intervariable & $37.1$ & $70.7$ & $89.7$\\
			\hline
		\end{tabular}
		\label{sigmathreshold}
	\end{table}
	
	Thus, the new formula to obtain predicted values for mass-loss rates based on the self-consistent wind solutions from \citet{alex19,alex21b} is:
	\begin{align}\label{mdotformula1}
		\log\dot M_\text{sc,predicted}=&-40.314 + 15.438\,w + 45.838\,x - 8.284\,w\,x \nonumber\\
 		&+ 1.0564\,y -  w\,y / 2.36 - 1.1967\,x\,y + 11.6\, z \nonumber\\
		&- 4.223\,w\,z - 16.377\,x\,z + y\,z / 81.735\;,
	\end{align}
	where $w$, $x$, $y$ and $z$ are defined as:
	\begin{equation}
		w=\log \left(\frac{T_\text{eff}}{\text{kK}}\right)\;,\nonumber\\
		x=\frac{1}{\log g}\;,\\
		y=\frac{R_*}{R_\odot}\;,\\
		z=\log \left(\frac{Z_*}{Z_\odot}\right)\;.
	\end{equation}
	
	Hence, this formula will be the basis of the self-consistent evolutionary tracks for the following sections of this paper.
	However, we include Appendix~\ref{extrafittings} where we introduce the parameters for the alternative linear and quadratic fittings, for illustrative purposes.
	
	One aspect that becomes clear from Eq.~\ref{mdotformula1} and remarks an important difference with previous studies, is that this intervariable fitting does not establish a rigid relationship between the mass-loss rate and the metallicity.
	Instead, we settle an interrelationship with the other stellar parameters.
	From \citet{vink01} it was determined that $\dot M\sim Z^{0.69\pm0.10}$ for O-type stars; whereas \citet{mokiem07} determined $\dot M\sim Z^{0.83\pm0.16}$ \citep[see review from][]{smith14} and, more recently, \citet{bjorklund21} determined $\dot M\sim Z^{0.95}$.
	In contrast, \citet{vink21} have provided a less steep relationship of $\dot M\sim Z^{0.42}$ for O-type stars.
	On the contrary, from Eq.~\ref{mdotformula1} and Tables~\ref{standardtable}, \ref{standardtablez07} and \ref{standardtablez03}, we derive that the scale between mass-loss rate and metallicity is easily correlated with the stellar mass.
	So, for a star born with 120 $M_\odot$ the relationship is $\dot M\sim Z^{0.53\pm0.01}$ \citep[an exponent far below the $\sim0.85$ found by previous authors with the exception of][]{vink21}, but for stars of 25 $M_\odot$ the relationship scales up to $\dot M\sim Z^{1.02\pm0.05}$.
	Intermediate exponents are then found for the stars with 70 $M_\odot$ ($\dot M\sim Z^{0.61\pm0.02}$) and with 40 $M_\odot$ ($\dot M\sim Z^{0.81\pm0.05}$; i.e., closer to those found by previous authors).
	Such deviation on the exponents for the relationship between mass-loss rate and metallicities demonstrates that this is not unique for all stars, and that stellar mass must be considered.
	
	Then, we incorporate Eq.~\ref{mdotformula1} as an extra option for the treatment of mass-loss rate in \textsc{Genec}, for $T_\text{eff}\ge30$ kK and $\log g\ge3.2$.
	Below each one of these values, mass-loss recipe is switched to the formula from \citet{vink01}.
	Effects over the evolutionary tracks of massive stars and their main implications are discussed in Section~\ref{evolutionarytracks}.

	\begin{table*}[t!]
		\centering
		\caption{\small{Self-consistent line-force parameters $(k,\alpha,\delta)$ for stars from \texttt{05} to \texttt{09} of the classical track (see Table~\ref{standardtable}), given the individual modification of the surface abundances as initial conditions. Last row of each group represents the combination of all the individual modifications previously outlined.}}
		\begin{tabular}{cc|ccc|cc}
			\hline
			\hline
			Name & modification in surface abundances & $k$ & $\alpha$ & $\delta$ & $\varv_\infty$ & $\log\dot M_\text{sc}$\\
			& & & & [km s$^{-1}$] & [$M_\odot\,\text{yr}^{-1}$]\\
			\hline
			\texttt{P120z10-05} & no modifications & 0.084 & 0.604 & 0.027 & $2\,730\pm140$ & $-5.354\pm.068$\\
			& [C/C$_\odot$]=0.5 & 0.084 & 0.604 & 0.027 & $2\,740\pm140$ & $-5.356\pm.068$\\
			& [N/N$_\odot$]=3.0 & 0.083 & 0.610 & 0.029 & $2\,780\pm140$ & $-5.335\pm.068$\\
			& total mod. in abundances & 0.083 & 0.609 & 0.028 & $2\,780\pm140$ & $-5.340\pm.068$\\
			\hline
			\texttt{P120z10-06} & no modifications & 0.077 & 0.635 & 0.041 & $2\,560\pm130$ & $-5.304\pm.065$\\
			& [He/H]=0.095 & 0.077 & 0.632 & 0.041 & $2\,560\pm130$ & $-5.309\pm.065$\\
			& [C/C$_\odot$]=0.03 & 0.077 & 0.631 & 0.039 & $2\,550\pm130$ & $-5.312\pm.065$\\
			& [N/N$_\odot$]=9.2 & 0.074 & 0.640 & 0.041 & $2\,610\pm130$ & $-5.306\pm.065$\\
			& [O/O$_\odot$]=0.45 & 0.076 & 0.620 & 0.032 & $2\,490\pm130$ & $-5.372\pm.065$\\
			& total mod. in abundances & 0.075 & 0.618 & 0.029 & $2\,510\pm130$ & $-5.409\pm.065$\\
			\hline
			\texttt{P120z10-07} & no modifications & 0.069 & 0.670 & 0.054 & $2\,510\pm200$ & $-5.067\pm.068$\\
			& [He/H]=0.11 & 0.068 & 0.667 & 0.055 & $2\,520\pm200$ & $-5.088\pm.068$\\
			& [C/C$_\odot$]=0.04 & 0.070 & 0.667 & 0.052 & $2\,510\pm200$ & $-5.076\pm.068$\\
			& [N/N$_\odot$]=11.0 & 0.068 & 0.677 & 0.056 & $2\,570\pm200$ & $-5.047\pm.068$\\
			& [O/O$_\odot$]=0.17 & 0.069 & 0.647 & 0.043 & $2\,370\pm190$ & $-5.162\pm.068$\\
			& total mod. in abundances & 0.067 & 0.644 & 0.040 & $2\,420\pm190$ & $-5.211\pm.068$\\
			\hline
			\texttt{P120z10-08} & no modifications & 0.061 & 0.681 & 0.063 & $2\,340\pm220$ & $-5.093\pm.100$\\
			& [He/H]=0.12 & 0.062 & 0.679 & 0.063 & $2\,400\pm220$ & $-5.101\pm.100$\\
			& [C/C$_\odot$]=0.04 & 0.060 & 0.677 & 0.059 & $2\,310\pm220$ & $-5.109\pm.100$\\
			& [N/N$_\odot$]=12.0 & 0.060 & 0.688 & 0.065 & $2\,390\pm220$ & $-5.080\pm.100$\\
			& [O/O$_\odot$]=0.1 & 0.063 & 0.673 & 0.068 & $2\,210\pm200$ & $-5.083\pm.100$\\
			& total mod. in abundances & 0.062 & 0.670 & 0.062 & $2\,300\pm210$ & $-5.129\pm.100$\\
			\hline
			\texttt{P120z10-09} & no modifications & 0.054 & 0.689 & 0.083 & $2\,040\pm210$ & $-5.081\pm.113$\\
			& [He/H]=0.13 & 0.053 & 0.690 & 0.078 & $2\,190\pm210$ & $-5.117\pm.113$\\
			& [C/C$_\odot$]=0.04 & 0.054 & 0.684 & 0.075 & $2\,050\pm210$ & $-5.124\pm.113$\\
			& [N/N$_\odot$]=12.0 & 0.053 & 0.693 & 0.082 & $2\,100\pm210$ & $-5.079\pm.113$\\
			& [O/O$_\odot$]=0.06 & 0.059 & 0.695 & 0.109 & $1\,910\pm190$ & $-4.962\pm.115$\\
			& total mod. in abundances & 0.055 & 0.691 & 0.091 & $2\,100\pm210$ & $-5.079\pm.113$\\
			\hline
		\end{tabular}
		\label{tableabundances}
	\end{table*}

\subsection{Variance of element abundances}\label{P120z10}
	The search of new line-force parameters is not limited only for the variations on effective temperature, surface gravity and stellar radius, but also for the abundances.
	One of the big advantages of the m-CAK prescription is the versatility to modify the abundances of the individual elements that compose the stellar wind.
	Even when metallicity can be considered as constant through the entire evolutionary track, nucleosynthesis processes through the evolution implies switches in the abundances for some elements.
	In turn, changes in the He to H ratio or in the individual abundance of metal elements affects the resulting line-acceleration\footnote{\textsc{Genec} provides us the change of abundances both on the core of the star and its surface. However, for the calculation of line-acceleration we are interested only on the modification of abundances in the surface of the star.} and therefore affecting the final mass-loss rate.
	We will also study such effects produced by abundances, in order to incorporate them to the final expressions for $\dot M$.
	Because \textsc{Genec} does not provide us surface abundances for important metals such as the iron-group, our analysis of individual metal elements is constrained to CNO elements only.

	Nevertheless, from the classical evolutionary tracks tabulated in Table~\ref{trackseries}, the only one which exhibits changes on surface abundances for CNO elements and the He to H ratio is  the track \texttt{P120z10}, for solar-metallicity and initial mass 120 $M_\odot$.
	This is a direct consequence of being working with non-rotating models: reactions in the core abruptly modify the abundance structure at some point for only extreme high mass cases such as 120 $M_\odot$.
	On the contrary, evolutionary tracks considering rotation exhibit a gradual modification of surface abundances even for stars with masses below 25 $M_\odot$ \citep{ekstrom12}.
	For the low metallicity scenarios, the non-rotating tracks shows an even late break-up for the surface abundances \citep{georgy13,eggenberger21}.
	Due to this reason, our analysis of the changes of abundances over the self-consistent mass-loss rate will be limited to the case of the track \texttt{P120z10}, as it follows.

	\begin{figure}[t!]
		\centering
		\includegraphics[width=0.9\linewidth]{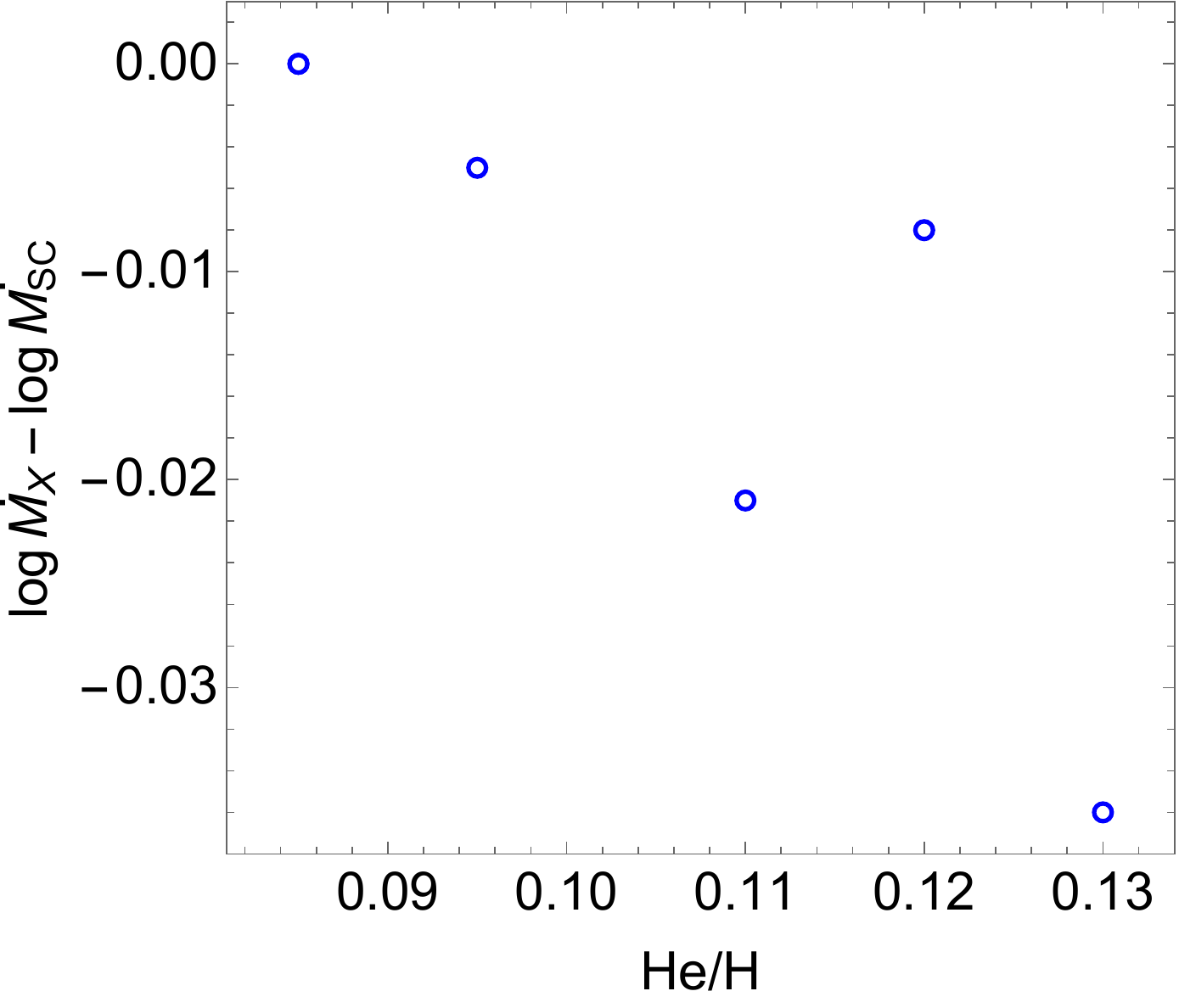}
		\caption{\small{Comparison between the resulting self-consistent mass-loss rate when the initial value for He to H ratio is modified, and the $\dot M_\text{sc}$ with the default value [He/H]=0.085.}}
		\label{abundmdot_heh}
	\end{figure}

	\begin{figure}[t!]
		\centering
		\includegraphics[width=0.9\linewidth]{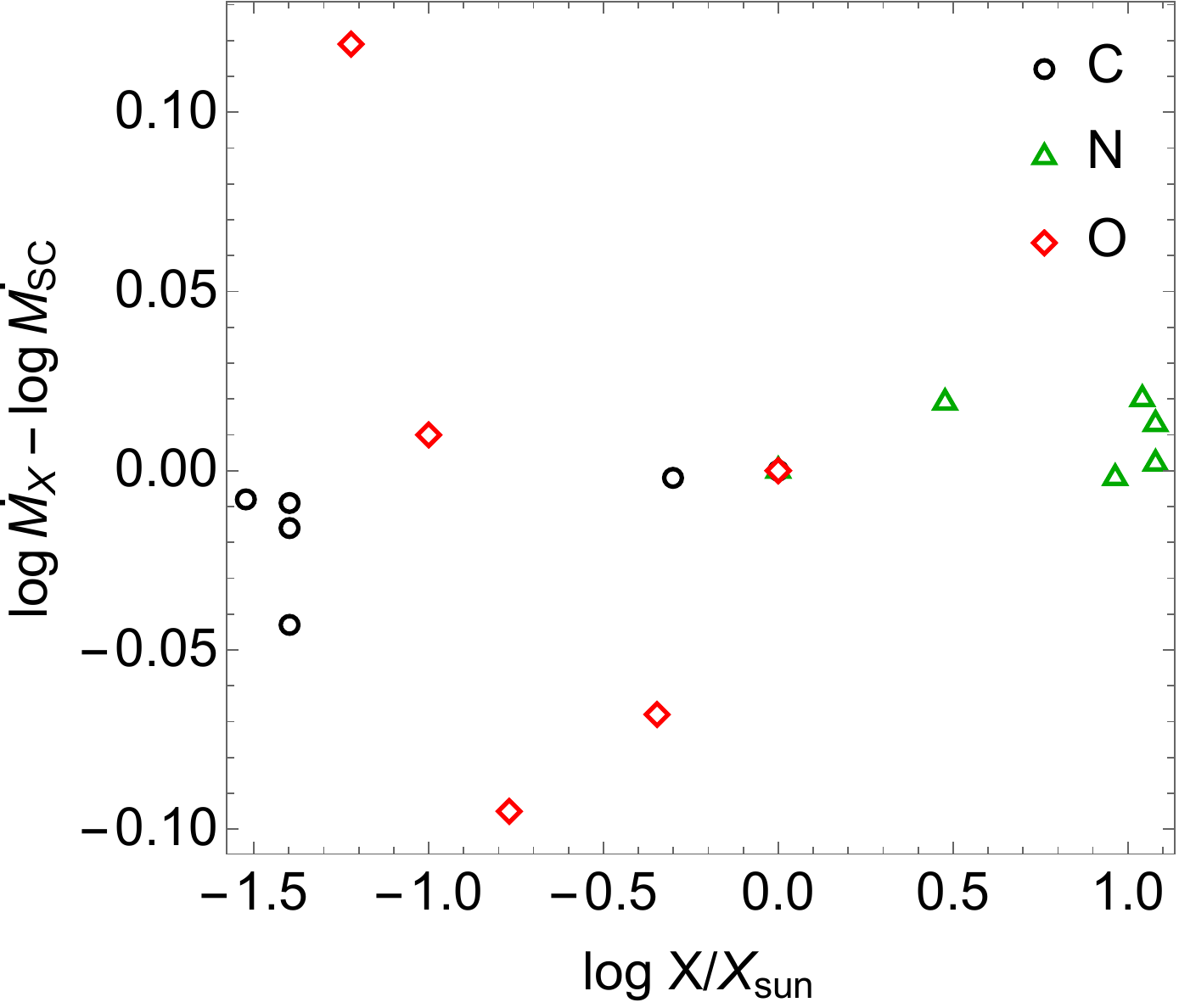}
		\caption{\small{Comparison between the resulting self-consistent mass-loss rate when the abundance for any of the CNO elements is modified, and the $\dot M_\text{sc}$ with the default solar value.}}
		\label{abundmdot_cno}
	\end{figure}

	We extract the stellar models from the track \texttt{P120z10}, specifically the models from \texttt{-05} to \texttt{-09}, because these are the only standard stars with modifications in their abundances in the stellar surface.
	Then we recalculate their self-consistent $(k,\alpha,\delta)$, according with such modifications in surface abundances.
	Results are shown in Table~\ref{tableabundances}.
	We see that, per each one of these individual models, a wind solution for the line-force parameters is in turn performed per each found variation on the abundance and subsequently a new $\dot M_\text{sc}$ is calculated.
	The ratios between these new mass-loss rates (with modified individual abundances) and the $M_\text{sc}$ with default abundances are shown in Fig.~\ref{abundmdot_heh} for the He to H ratio and in Fig.~\ref{abundmdot_cno} for the CNO elements.

	These variations however, confirm that it is not possible to establish a clear dependence between the abundances and the mass-loss rate, especially if we consider that the error bars for the values of $\dot M_\text{sc}$ lie between $\pm0.06$ and $\pm0.11$.
	In other words, the erratic variances presented in Table ~\ref{tableabundances} and in Figures~\ref{abundmdot_heh} and \ref{abundmdot_cno} can be better explained as a product of the uncertainties carried from the self-consistent wind solution instead.
	Given this result, we decide not to include effects from variations in the surface CNO elements nor in the He-to-H ratio over the theoretical mass-loss rate.
	Such analysis will be reserved for a forthcoming study, dedicated to the case of $\dot M$ for rotating-models.

\section{Evolutionary models}\label{evolutionarytracks}	
	\begin{figure*}[t!]
		\centering
		\includegraphics[width=0.33\linewidth]{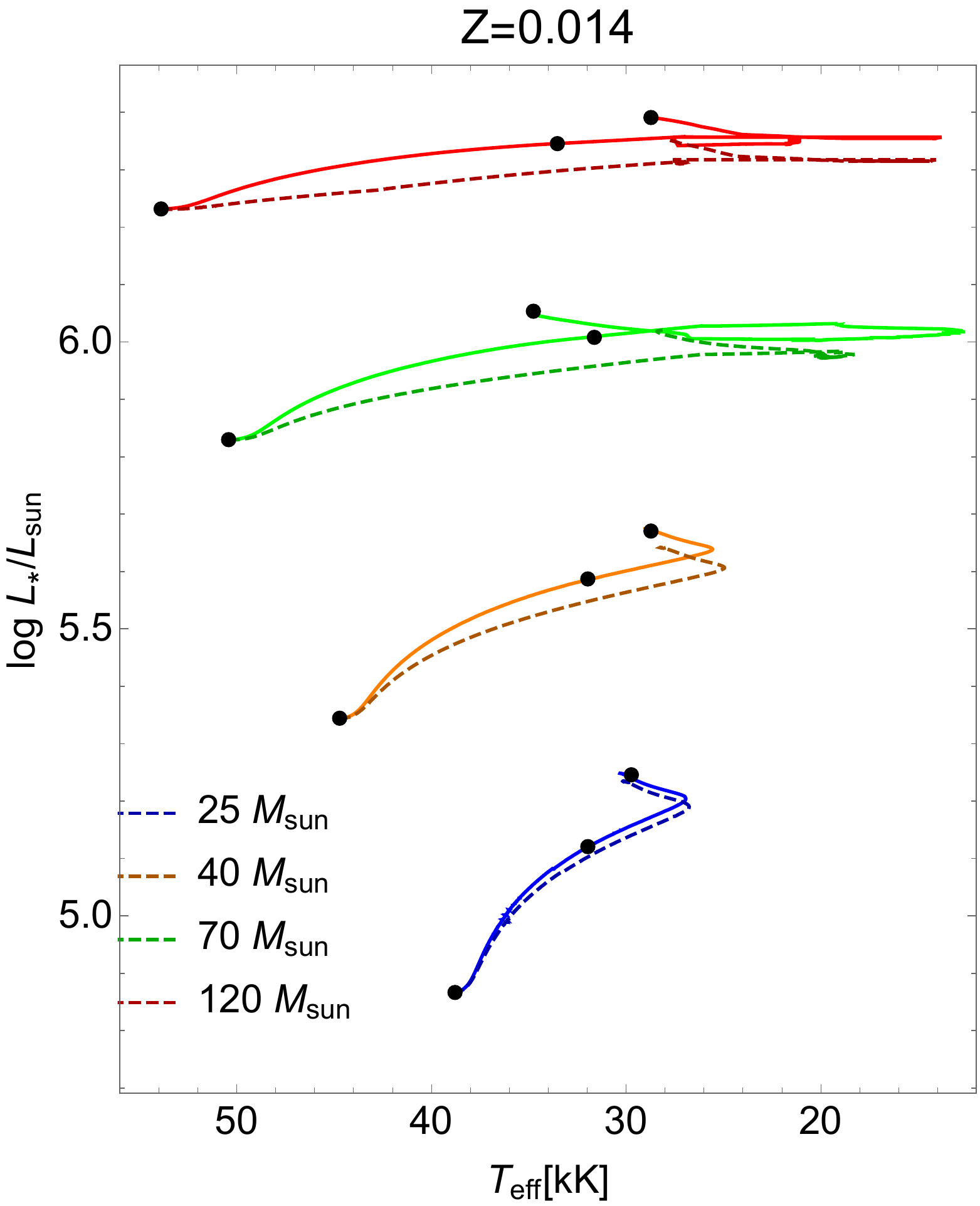}
		\includegraphics[width=0.33\linewidth]{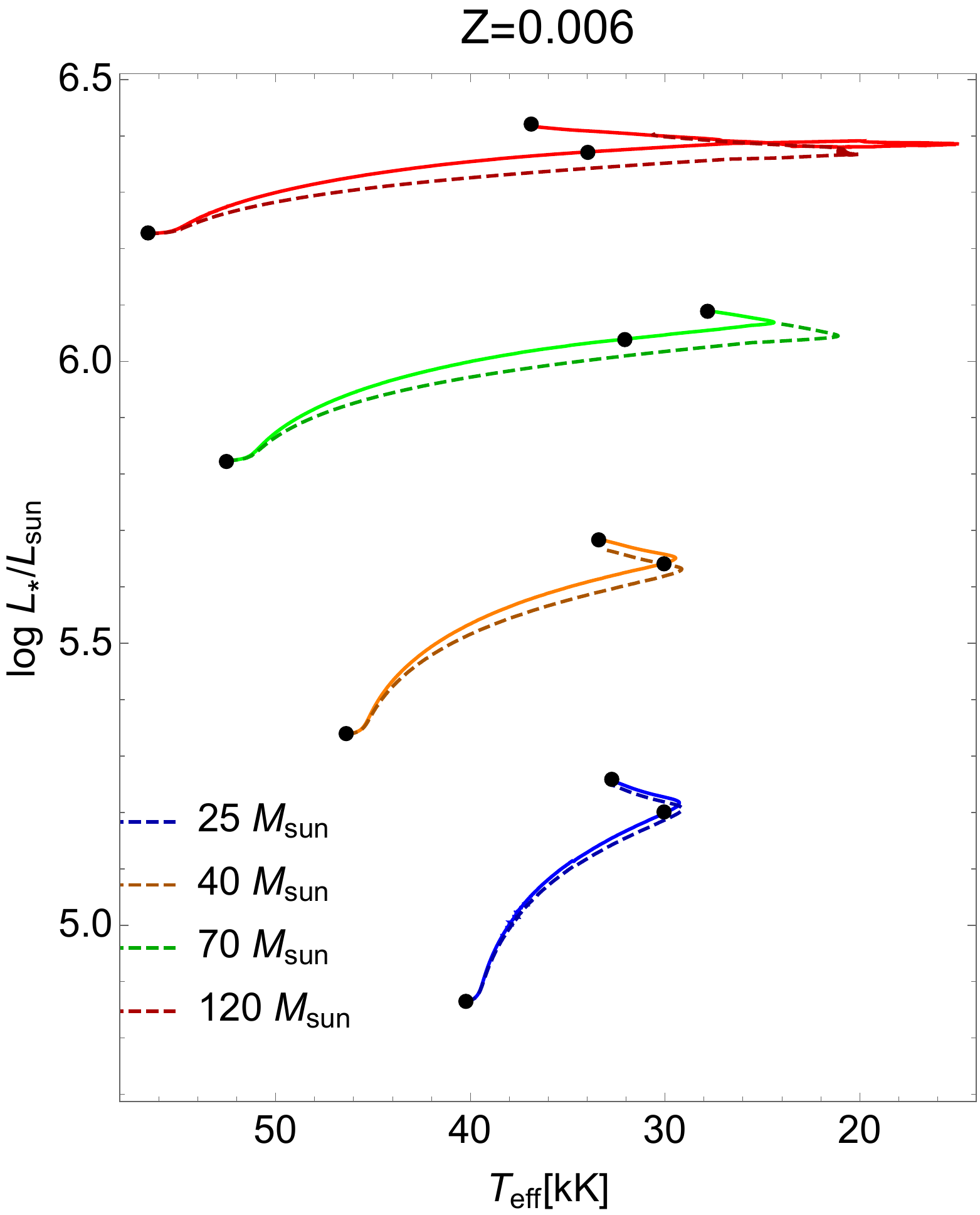}
		\includegraphics[width=0.33\linewidth]{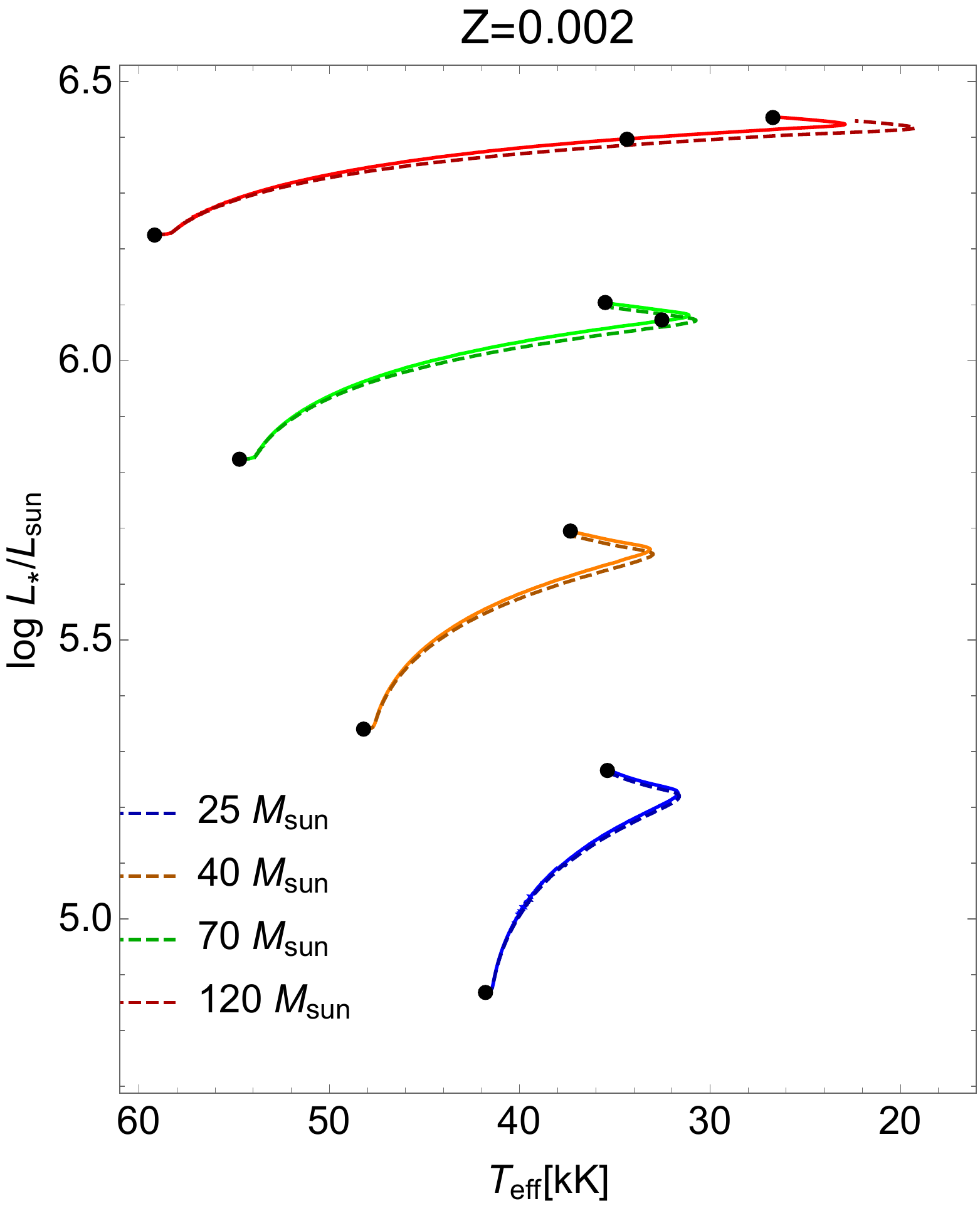}
		\caption{\small{Evolutionary tracks followed across the Hertzsprung-Russell diagram for model stars without rotation, assuming the mass-loss rates given originally from \citet[][dashed lines, i.e., classical tracks]{vink00} and those $\dot M_\text{sc}$ given by the self-consistent wind solutions from \citet[][continue lines, i.e., self-consistent tracks]{alex21b}.
		 Black dots represent: i) the ZAMS-point, ii) where $\dot M_\text{sc}$ switches to $\dot M_\text{Vink}$, and iii) the end of the H-burning stage.}}
		\label{HDR_final}
	\end{figure*}

\begin{figure*}[t!]
		\centering
		\includegraphics[width=0.33\linewidth]{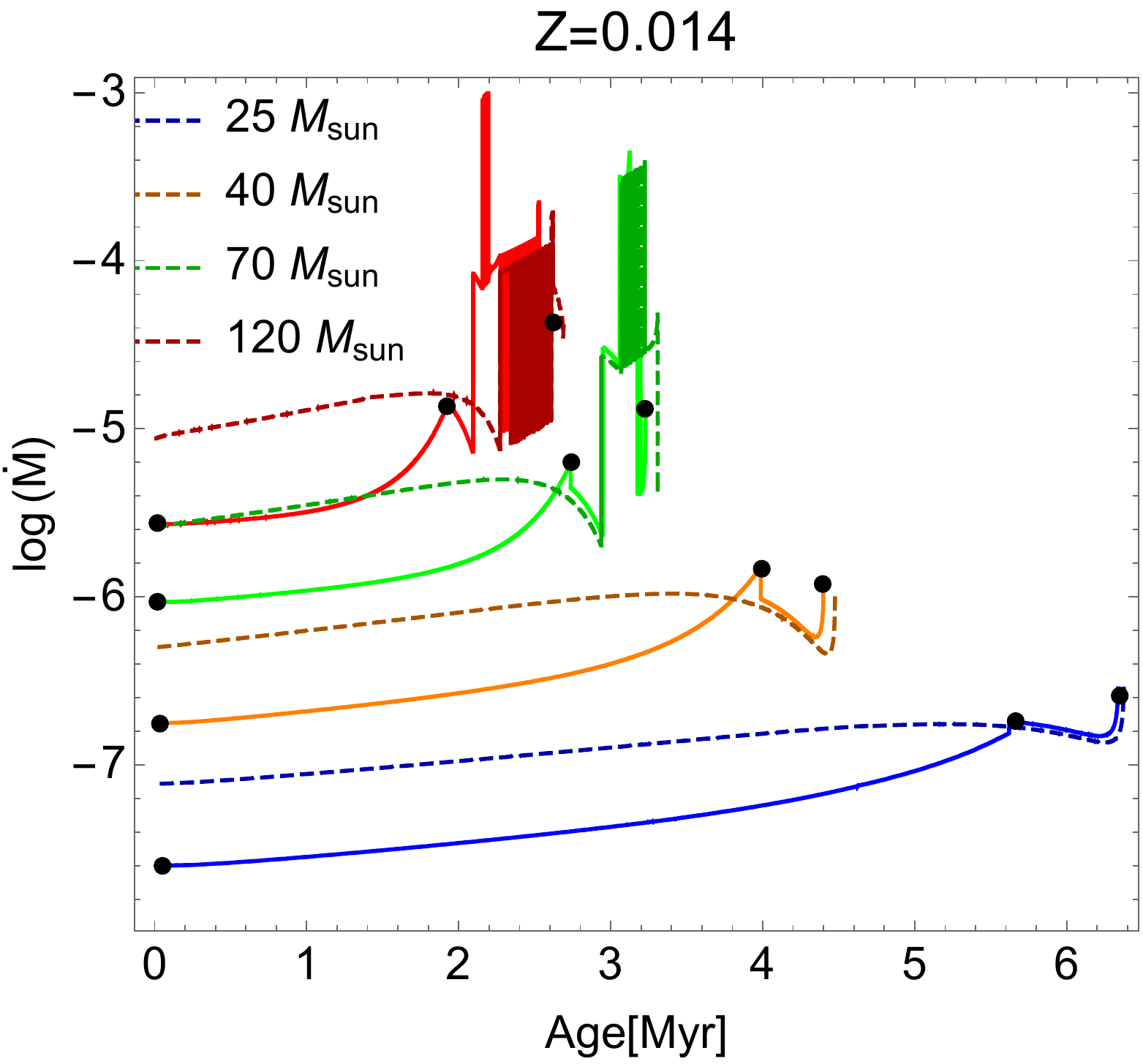}
		\includegraphics[width=0.33\linewidth]{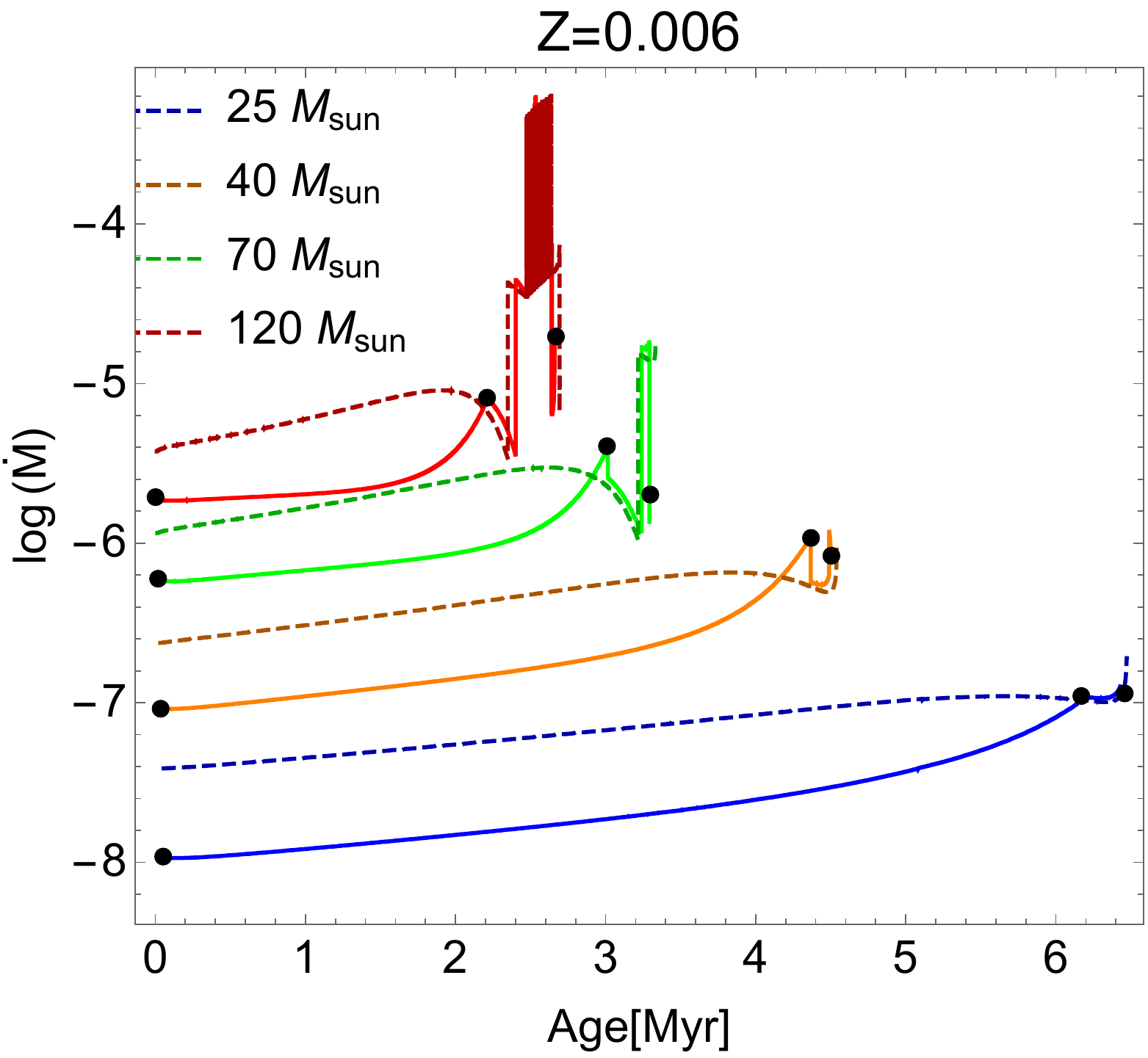}
		\includegraphics[width=0.33\linewidth]{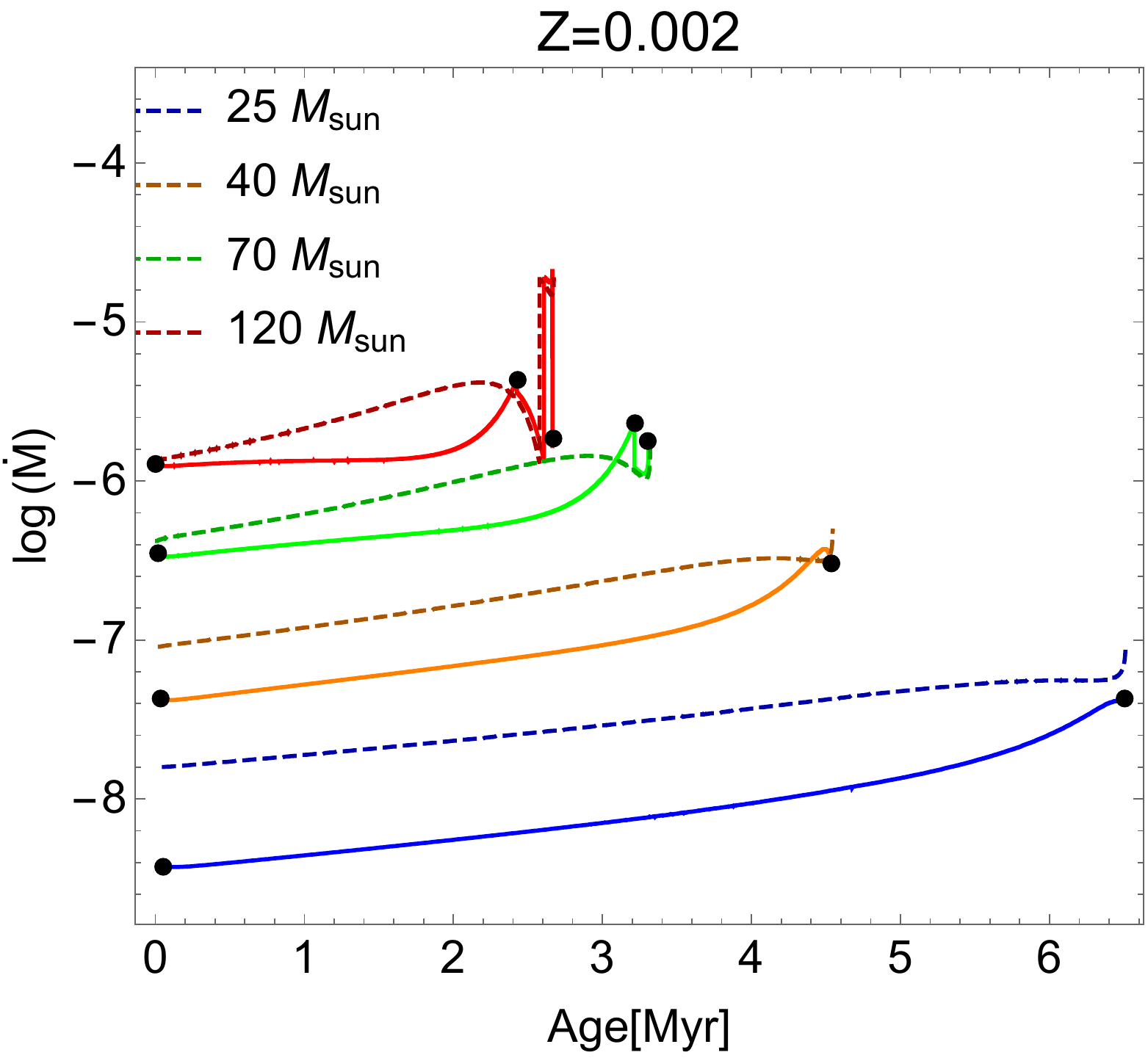}
		\caption{\small{Evolution of $\dot M$ of evolutionary tracks for stars with 120, 70, 40 and 25 $M_\odot$, calculated for self-consistent tracks (solid lines) and for classical tracks (dashed lines).
		Black dots represent the same as indicated in Fig.~\ref{HDR_final}.}}
		\label{mdot_fin}
	\end{figure*}

\begin{figure*}[t!]
		\centering
		\includegraphics[width=0.33\linewidth]{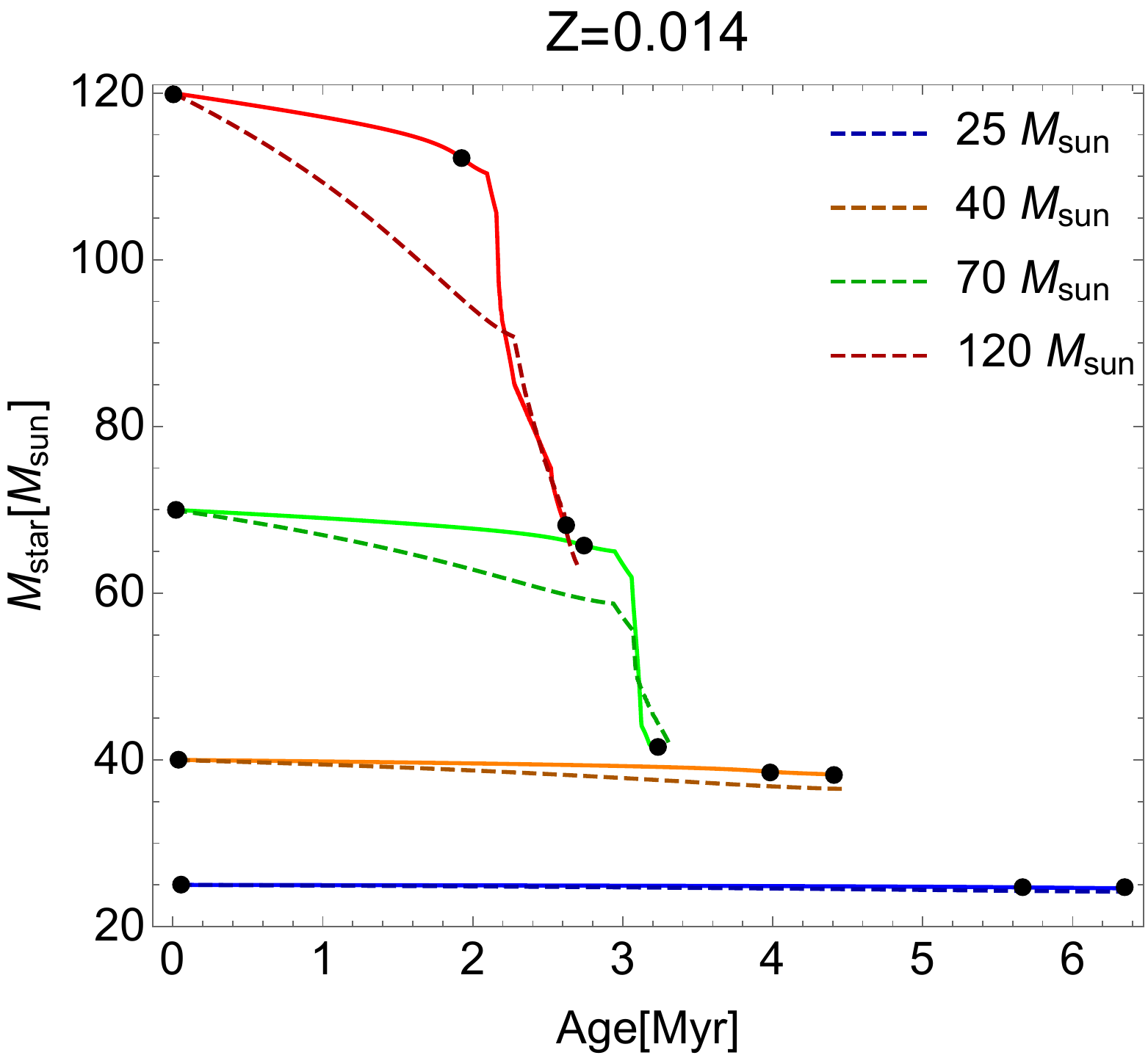}
		\includegraphics[width=0.33\linewidth]{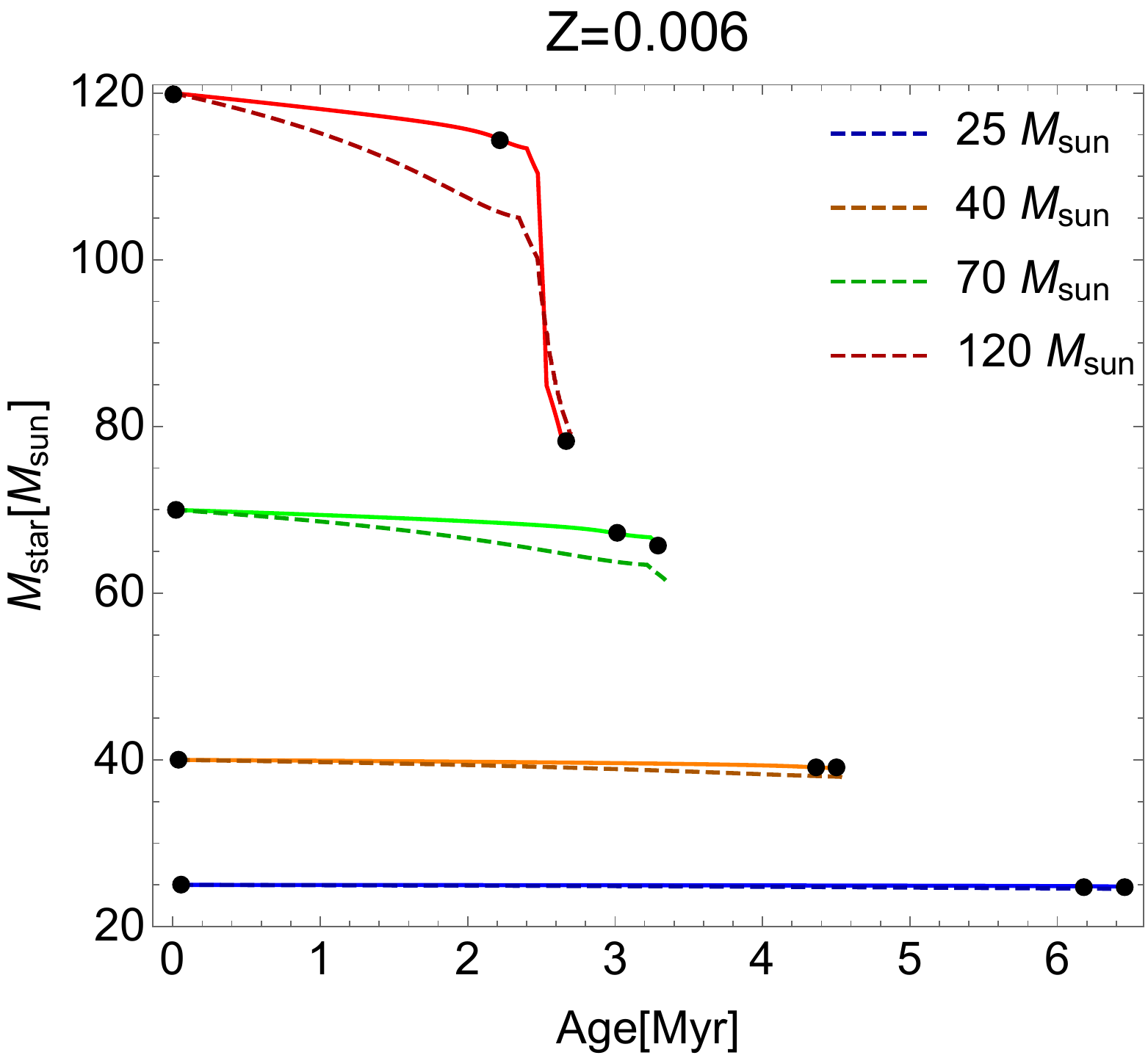}
		\includegraphics[width=0.33\linewidth]{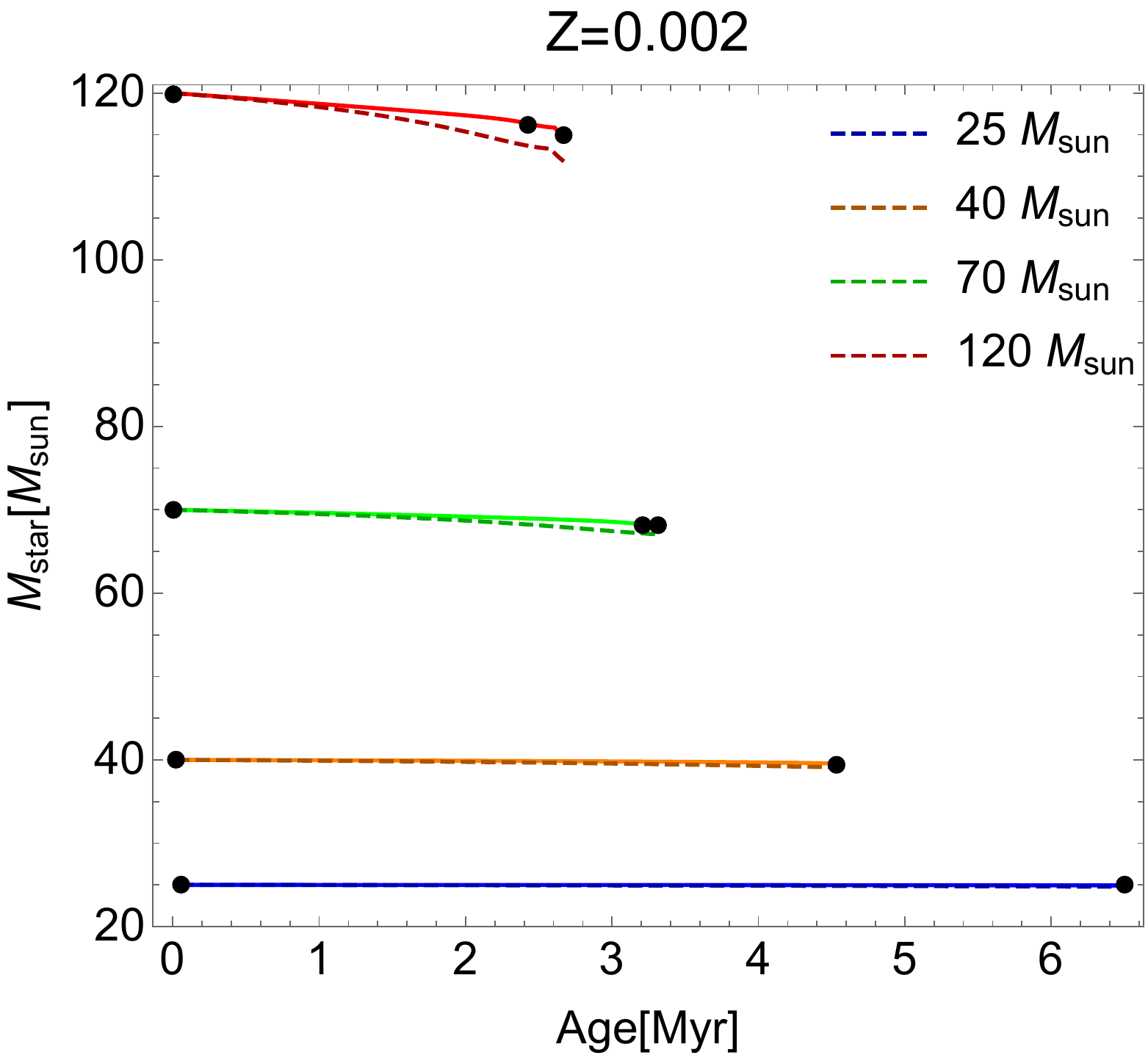}
		\caption{\small{Evolution of the stellar masses $M_*$ of our evolutionary tracks for stars with 120, 70, 40 and 25 $M_\odot$, calculated for self-consistent tracks (solid lines) and for classical tracks (dashed lines).
		 Black dots represent the same as indicated in Fig.~\ref{HDR_final}.}}
		\label{mass_fin}
	\end{figure*}

	HR diagrams, with the different evolutionary tracks, are shown in Fig.~\ref{HDR_final}.
	Evolution of mass-loss rates is shown in Fig.~\ref{mdot_fin}, whereas evolution of stellar masses appears in Fig.~\ref{mass_fin}.
	For these three sets of figures, the evolutionary tracks cover from the ZAMS to the end of the H-burning stage.
	Besides, for each one of these figures, we included three black dots which corresponds to: i) the ZAMS-point, ii) the point where the self-consistent recipe for the mass-loss rate ends and iii) the point where we reach the end of the H-burning stage, as we will see in the following Subsections.
	We use these points as tracers to explore the new stellar and wind structure through the different evolutionary tracks.
	Extra properties of the tracks, such as lifetimes and final surface abundances, are summarised in Table~\ref{timescalesXY}.
	Hereafter, evolutionary tracks using $\dot M_\text{Vink}$ will be denoted again as \textit{classical tracks}\footnote{Note that classical tracks introduced in Section~\ref{method} (Tables~\ref{standardtable}, \ref{standardtablez07} and \ref{standardtablez03}) are not the same as tracks from this Section~\ref{evolutionarytracks}, which were performed to extract standard stellar models and calculate self-consistent wind solutions.}, whereas tracks using $\dot M_\text{sc}$ will be called \textit{self-consistent tracks}.
	
\begin{figure*}[t!]
		\centering
		\includegraphics[width=0.33\linewidth]{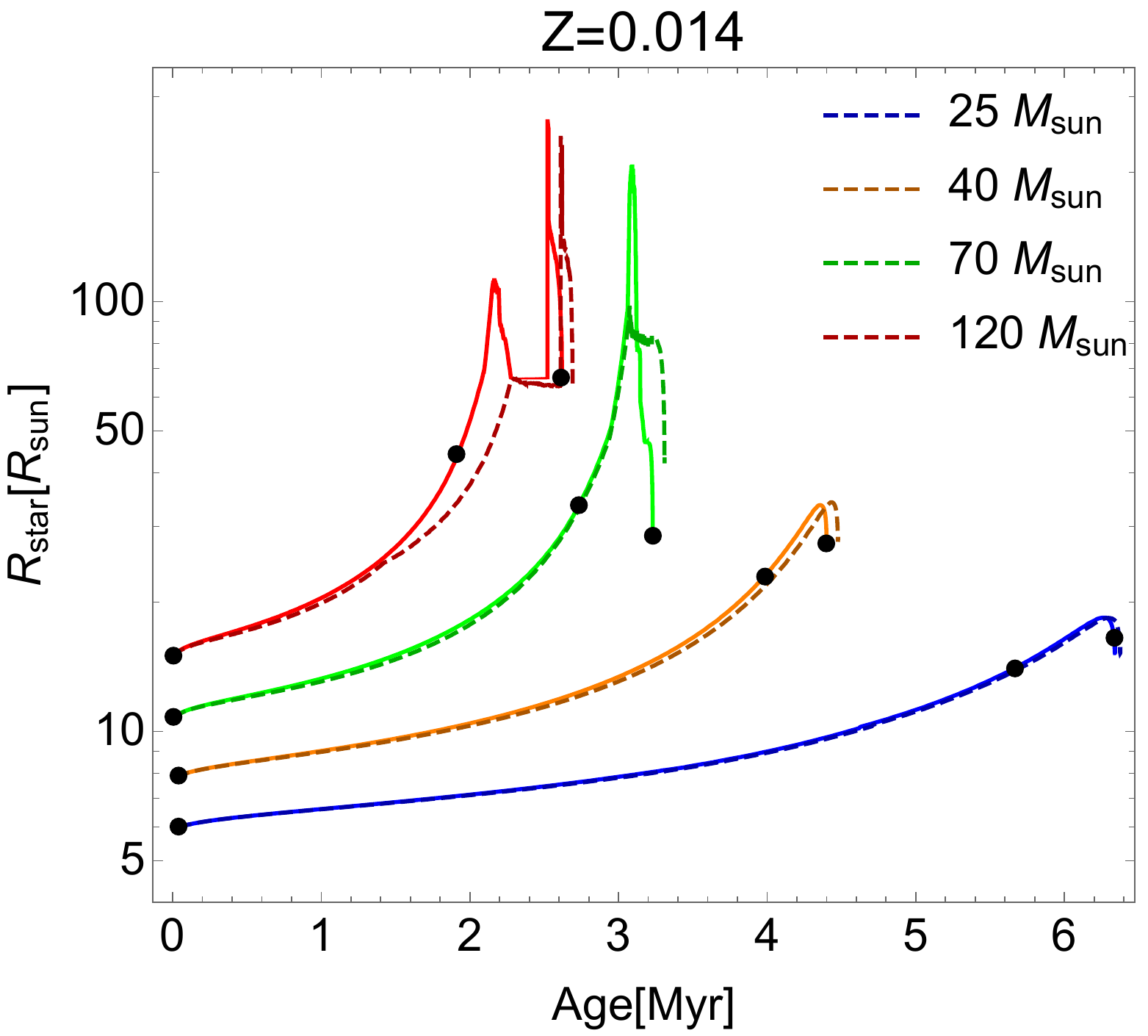}
		\includegraphics[width=0.33\linewidth]{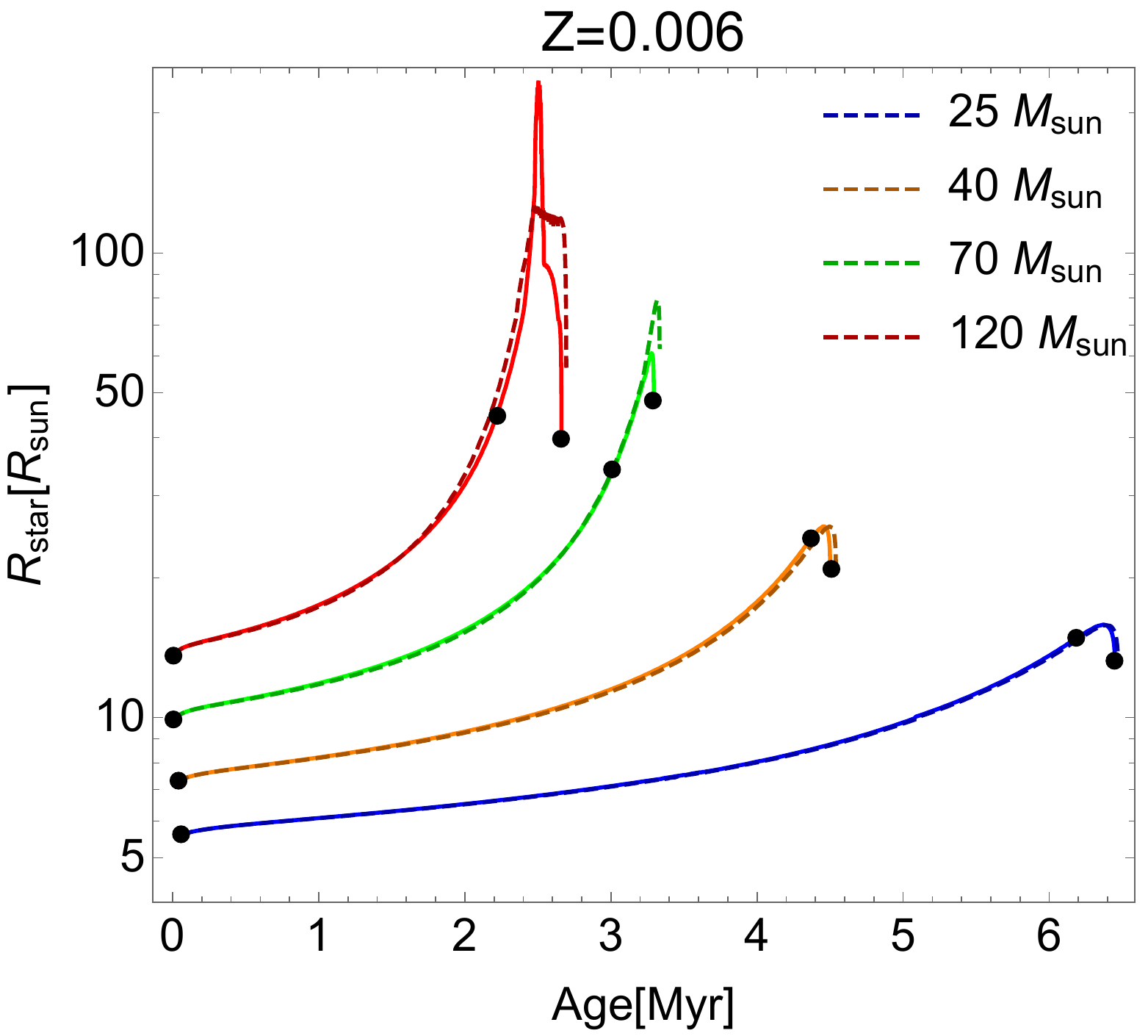}
		\includegraphics[width=0.33\linewidth]{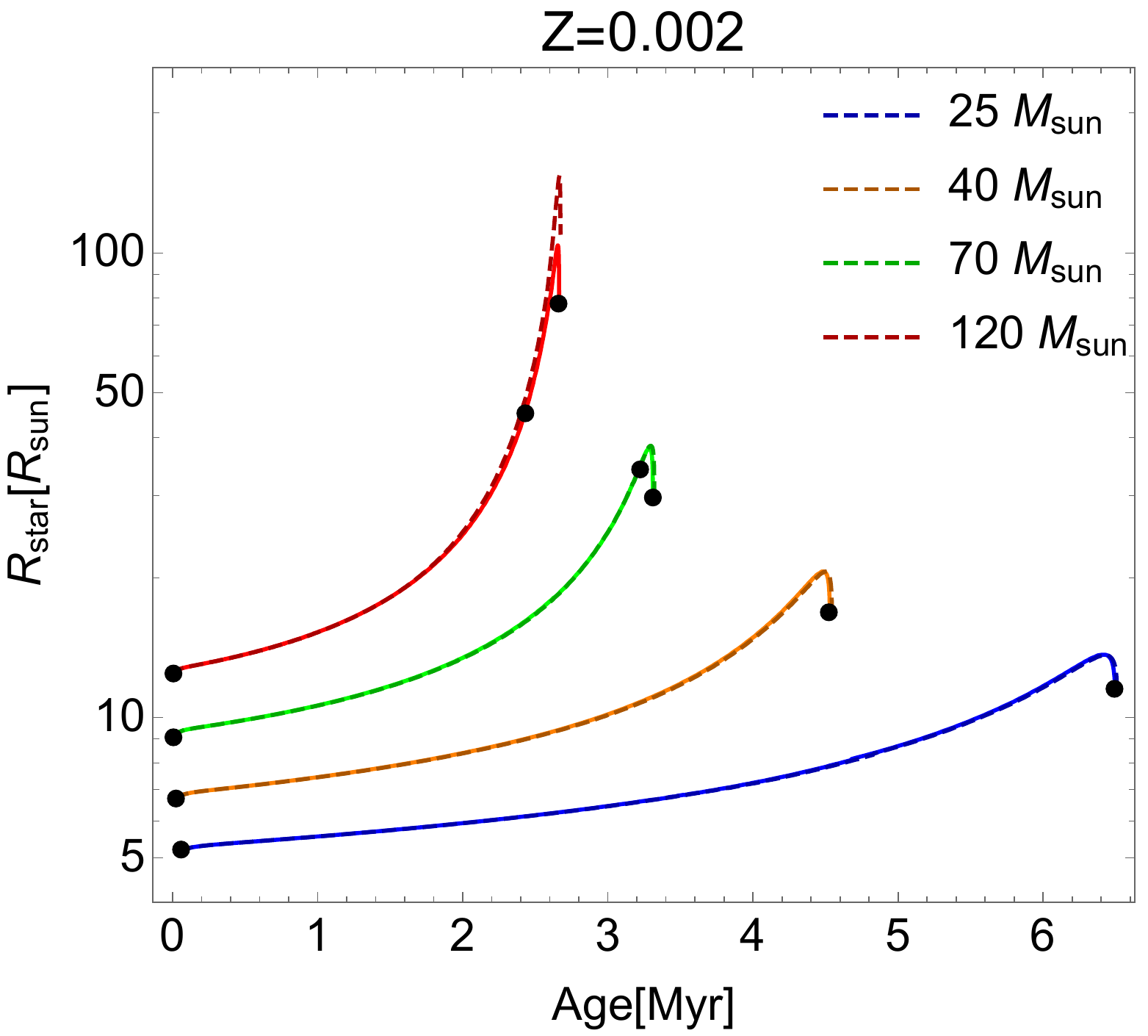}
		\caption{\small{Evolution of the stellar radii $R_*$ of our evolutionary tracks for stars with 120, 70, 40 and 25 $M_\odot$, calculated for self-consistent tracks (solid lines) and for classical tracks (dashed lines).
		Black dots represent the same as indicated in Fig.~\ref{HDR_final}.}}
		\label{radii_fin}
	\end{figure*}

\begin{figure*}[t!]
		\centering
		\includegraphics[width=0.33\linewidth]{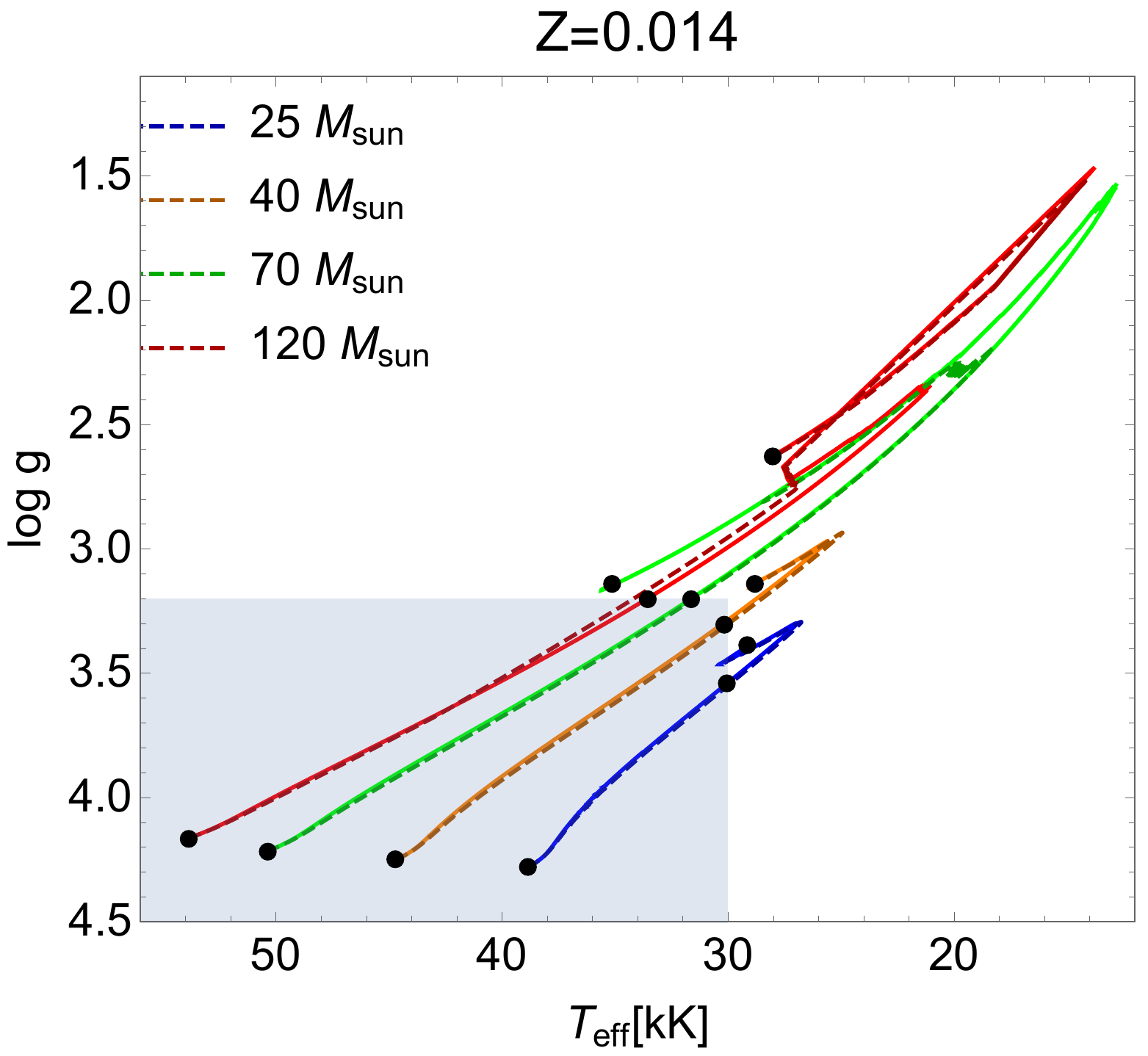}
		\includegraphics[width=0.33\linewidth]{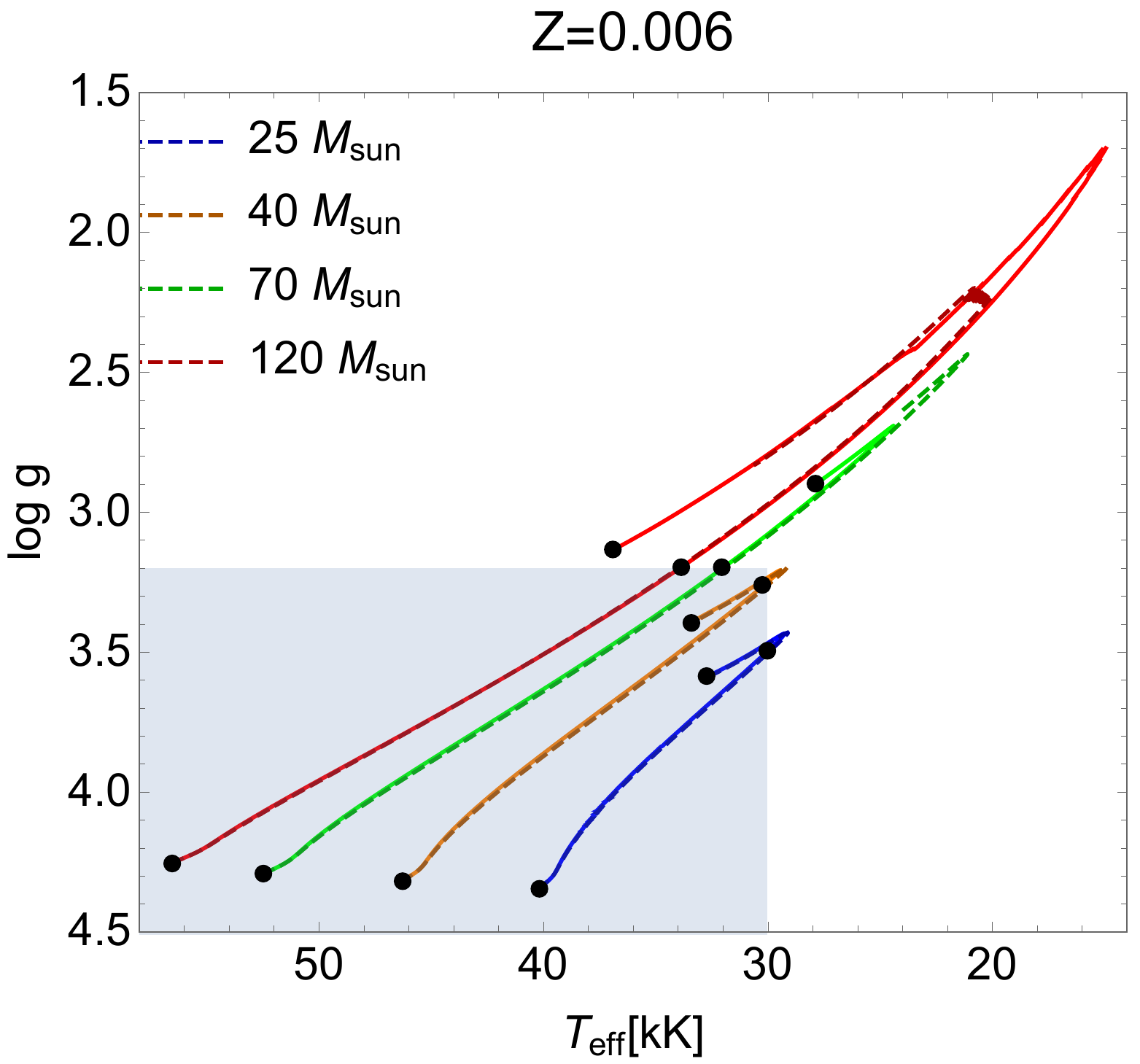}
		\includegraphics[width=0.33\linewidth]{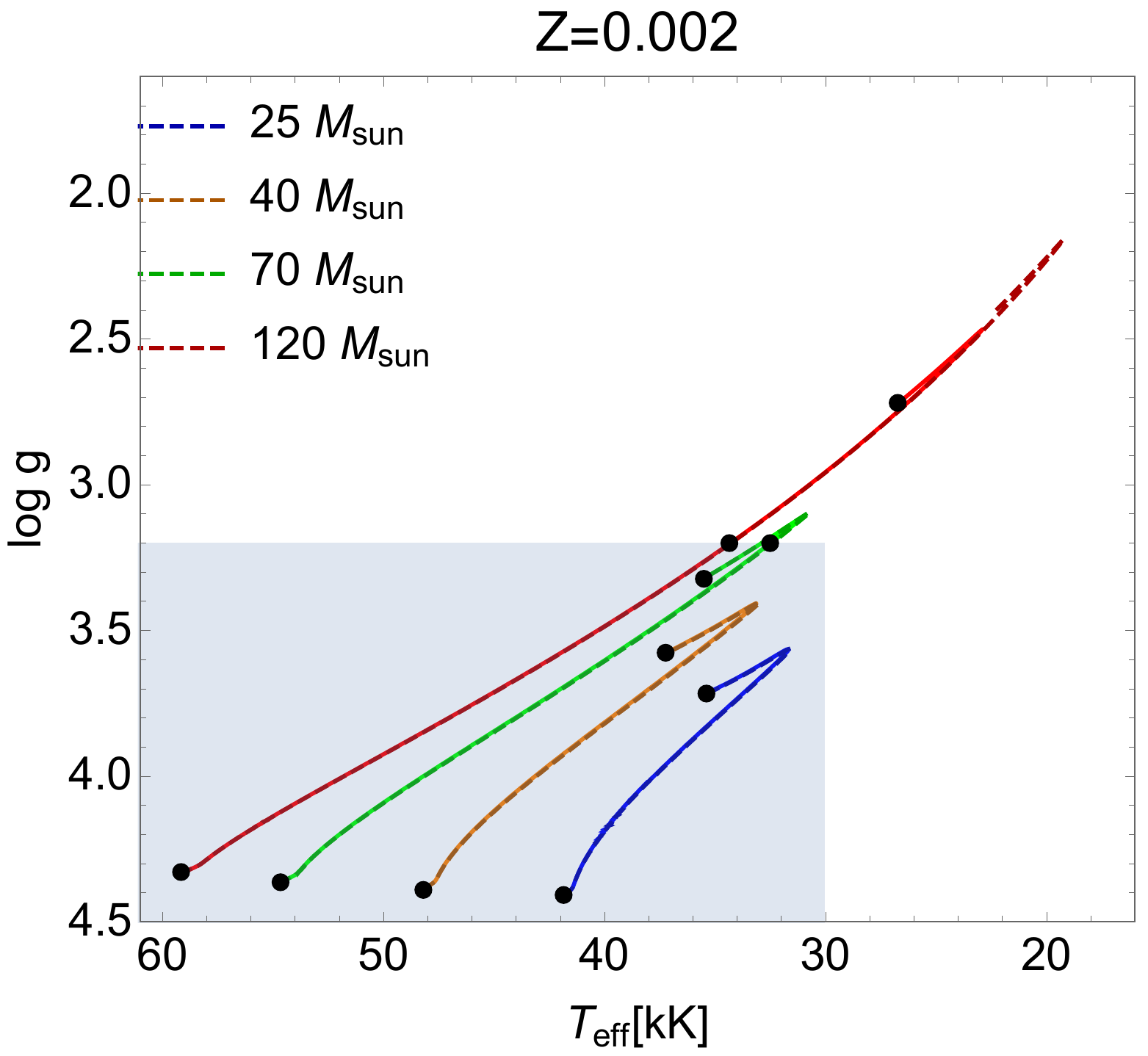}
		\caption{\small{Evolutionary tracks across the spectroscopic Hertzsprung-Russell diagram \citep{langer14} for model stars without rotation, calculated for self-consistent tracks (solid lines) and for classical tracks (dashed lines). Grey shaded area represents the region where self-consistent m-CAK prescription is valid \citep[$T_\text{eff}\ge30$, $\log g\ge3.2$, see][]{alex21b}.
		Black dots represent the same as indicated in Fig.~\ref{HDR_final}.}}
		\label{sHDR_final}
	\end{figure*}

	The most remarkable result, from Fig.~\ref{HDR_final}, is that self-consistent evolutionary tracks are drifted towards more luminous scales in the HD diagrams.
	Such movements seem to be a direct consequence of $\dot M_\text{sc}$ being less strong than $\dot M_\text{Vink}$, as shown in Fig.~\ref{mdot_fin}.
	Stars evolving within self-consistent tracks retain more mass than within classical tracks (Fig.~\ref{mass_fin}), resulting into both more massive and larger stars in terms of stellar radii (Fig.~\ref{radii_fin}) and more luminous stars in turn.
	The direct relation between the mass and the radius of the stars becomes more evident if we plot the so-called spectroscopic HR diagram \citep[sHRD,][]{langer14}, where we observe that there is no significative difference between the surface gravities, proportional to $M_*/R_*^2$, from the different tracks (Fig.~\ref{sHDR_final}).
	
	This departure between the projected tracks is more prominent for higher mass and metallicities; which can be attributed to the bigger difference in mass-loss rate between self-consistent and Vink's formula for these case, but also to the absolute quantity of the mass-loss rate, as seen in Fig.~\ref{mdot_fin}.
	Moreover, this is also in agreement with the level of relevance that mass-loss has over the evolution of O-type stars: from full dominance for stars with $M\gtrsim60$ $M_\odot$ \citep{vink12}; a slightly less relevant role for stars $30M_\odot<M_*<60M_\odot$ \citep{langer12,groh14}; and a secondary influence for stars with $M_*<30M_\odot$, where features such as rotation (absent in our study) shows a preponderance \citep{maeder00}.	

	\begin{table*}[htbp]
		\centering
		\caption{\small{Properties of the stellar models at the end of the H-burning phases..}}
		\begin{tabular}{ccc|cccccc}
			\hline
			\hline
			$M_\text{ZAMS}$ & $Z$ & Mass-loss & \multicolumn{6}{c}{End of H-burning}\\
			& & & $t_\text{H}$ & $M_*$ & $R_*$ & $Y_\text{surf}$ & N/C & N/O\\
			\hline
			120.0 & 0.014 & $\dot M_\text{Vink}$ & 2.672 & 63.625 & 63.7 & 0.783 & 84.87 & 67.54 \\ 
			& & $\dot M_\text{sc}$ & 2.619 & 68.076 & 66.1 & 0.800 & 82.76 & 68.67 \\ 
			120.0 & 0.006 & $\dot M_\text{Vink}$ & 2.695 & 79.146 & 56.7 & 0.588 & 94.05 & 68.29 \\ 
			& & $\dot M_\text{sc}$ & 2.661 & 78.189 & 39.8 & 0.641 & 103.9 & 69.84 \\ 
			120.0 & 0.002 & $\dot M_\text{Vink}$ & 2.674 & 111.76 & 110. & 0.251 & 0.332 & 0.101 \\ 
			& & $\dot M_\text{sc}$ & 2.664 & 114.82 & 77.4 & 0.251 & 0.332 & 0.101 \\ 
			\hline
			70.0 & 0.014 & $\dot M_\text{Vink}$ & 3.310 & 41.933 & 42.9 & 0.543 & 133.0 & 60.16 \\ 
			& & $\dot M_\text{sc}$ & 3.323 & 41.519 & 28.7 & 0.617 & 127.1 & 60.02 \\ 
			70.0 & 0.006 & $\dot M_\text{Vink}$ & 3.334 & 62.895 & 61.7 & 0.256 & 0.332 & 0.101 \\ 
			& & $\dot M_\text{sc}$ & 3.295 & 47.802 & 65.8 & 0.256 & 0.332 & 0.101 \\ 
			70.0 & 0.002 & $\dot M_\text{Vink}$ & 3.322 & 67.051 & 30.1 & 0.251 & 0.332 & 0.101 \\ 
			& & $\dot M_\text{sc}$ & 3.311 & 68.119 & 29.8 & 0.251 & 0.332 & 0.101 \\ 
			\hline
			40.0 & 0.014 & $\dot M_\text{Vink}$ & 4.471 & 36.508 & 28.1 & 0.266 & 0.332 & 0.101 \\ 
			& & $\dot M_\text{sc}$ & 4.401& 38.253 & 27.6 & 0.266 & 0.332 & 0.101 \\ 
			40.0 & 0.006 & $\dot M_\text{Vink}$ & 4.541 & 37.966 & 20.8 & 0.256 & 0.332 & 0.101 \\ 
			& & $\dot M_\text{sc}$ & 4.501 & 39.007 & 20.8 & 0.256 & 0.332 & 0.101 \\ 
			40.0 & 0.002 & $\dot M_\text{Vink}$ & 4.546 & 39.102 & 16.8 & 0.251 & 0.332 & 0.101 \\ 
			& & $\dot M_\text{sc}$ & 4.531 & 39.559 & 16.9 & 0.251 & 0.332 & 0.101 \\ 
			\hline
			25.0 & 0.014 & $\dot M_\text{Vink}$ & 6.376 & 24.178 & 15.5 & 0.266 & 0.332 & 0.101 \\ 
			& & $\dot M_\text{sc}$ & 6.342 & 24.591 & 16.6 & 0.266 & 0.332 & 0.101 \\ 
			25.0 & 0.006 & $\dot M_\text{Vink}$ & 6.473 & 24.520 & 13.1 & 0.256 & 0.332 & 0.101 \\ 
			& & $\dot M_\text{sc}$ & 6.452 & 24.808 & 13.2 & 0.256 & 0.332 & 0.101 \\ 
			25.0 & 0.002 & $\dot M_\text{Vink}$ & 6.512 & 24.779 & 11.4 & 0.251 & 0.332 & 0.101 \\ 
			& & $\dot M_\text{sc}$ & 6.498 & 24.931 & 11.5 & 0.251 & 0.332 & 0.101 \\ 
			\hline
		\end{tabular}
		\label{timescalesXY}
	\end{table*}

	\begin{figure*}[t!]
		\centering
		\includegraphics[width=0.33\linewidth]{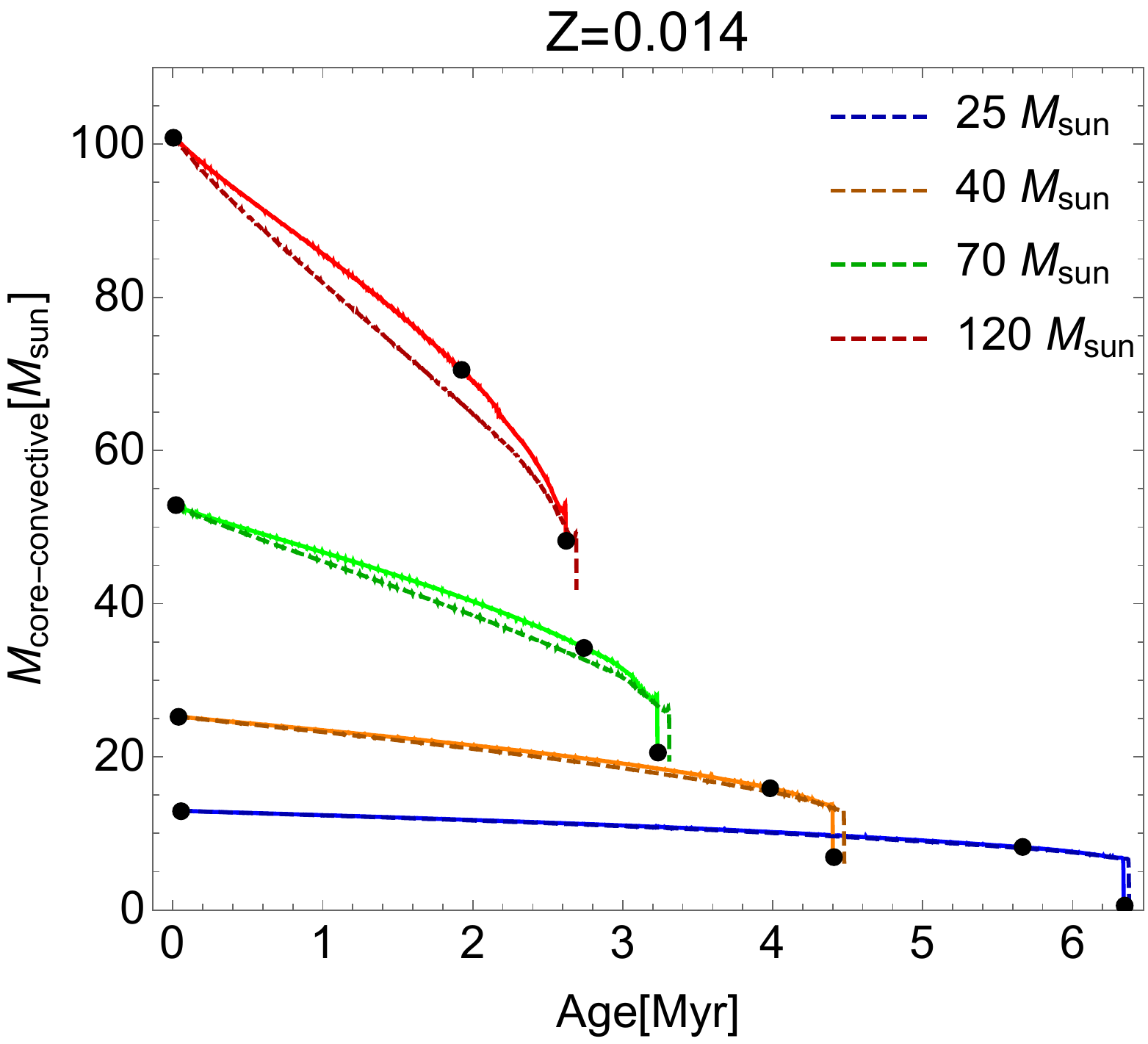}
		\includegraphics[width=0.33\linewidth]{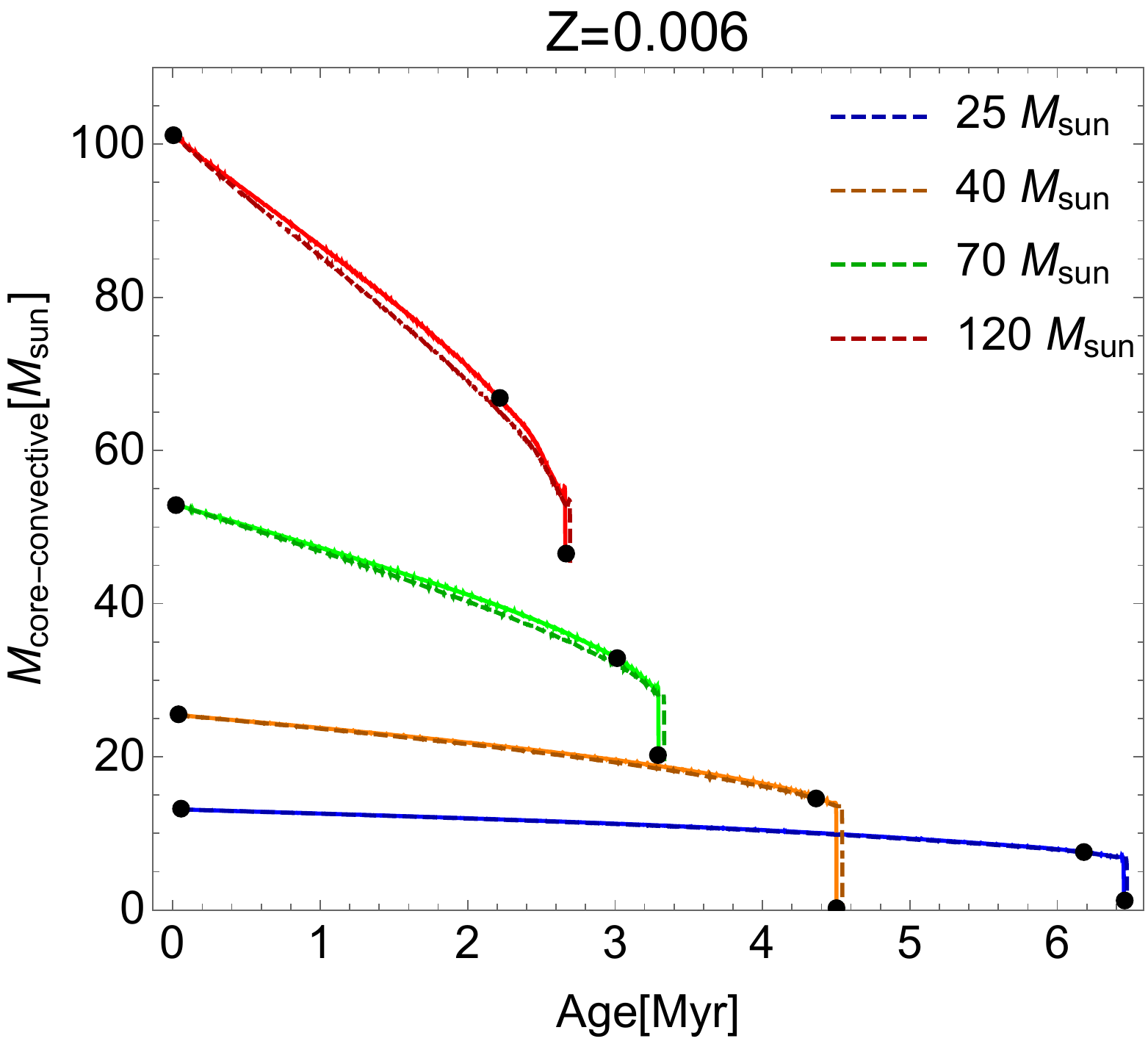}
		\includegraphics[width=0.33\linewidth]{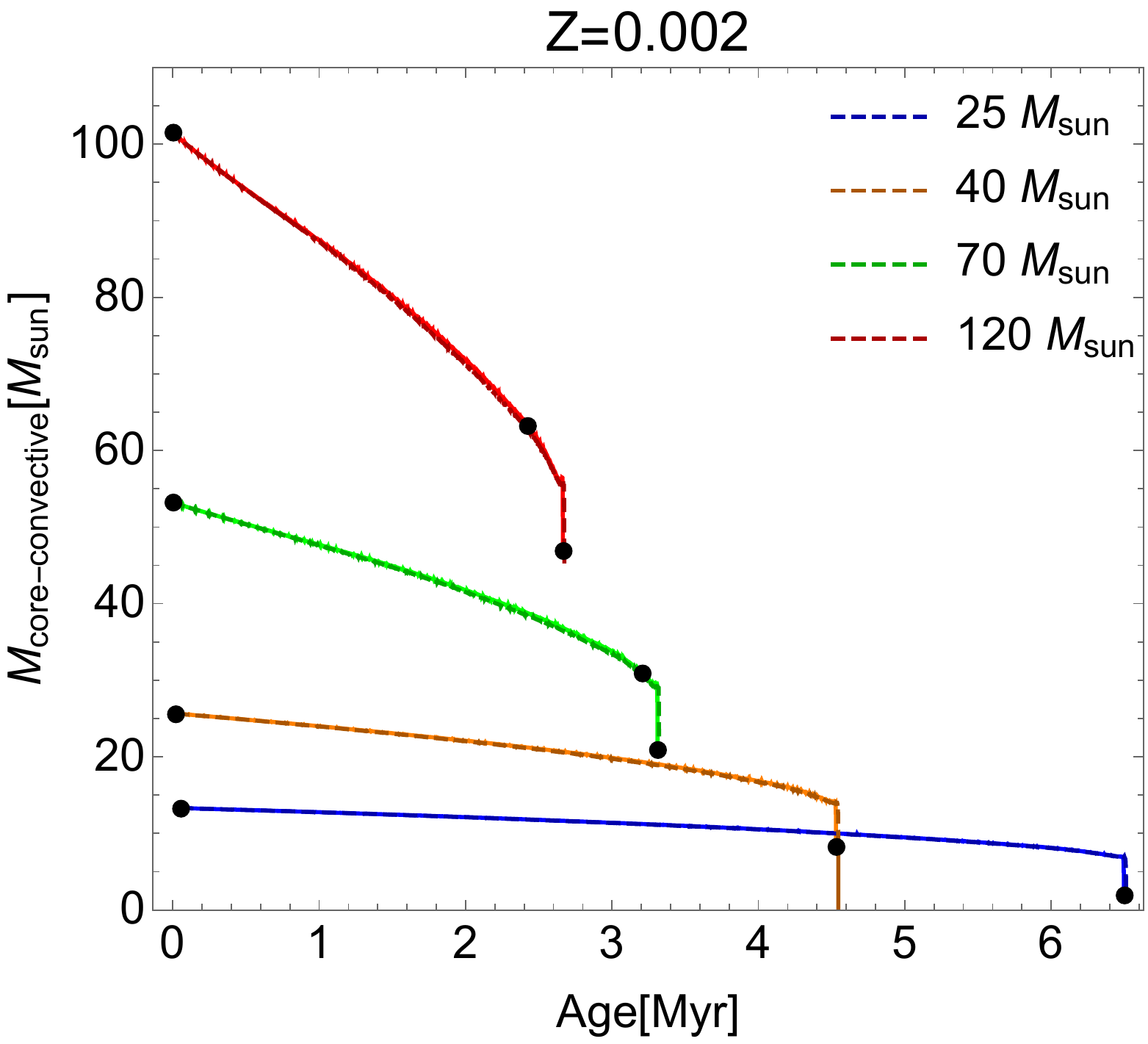}
		\caption{\small{Evolution of the mass of the convective-core $M_\text{cc}$ (relative to the total stellar mass $M_*$), of our evolutionary tracks for stars with 120, 70, 40 and 25 $M_\odot$, calculated using our self-consistent $\dot M_\text{sc}$ (solid lines) and using Vink's formula $\dot M_\text{Vink}$ (dashed lines).
		Black dots represent the same as indicated in Fig.~\ref{HDR_final}.}}
		\label{ccmass_fin}
	\end{figure*}
	
\subsection{Cases $M_\text{ZAMS}=120\;M_\odot$}
	As expected, the largest differences between the evolutionary tracks using our self-consistent prescription for mass-loss rate are given for the most massive cases.
	This is particularly evident from Fig.~\ref{mass_fin}, where the differences in the evolution of the stellar masses are more pronounced, especially for the more metallic cases.
	Stars from the tracks using $\dot M_\text{sc}$ lose their masses at a slower rate, retaining $\sim20$ $M_\odot$ of extra mass for the Galactic case ($Z=0.014$), $\sim10$ $M_\odot$ for the LMC ($Z=0.006$) case and $\sim5$ $M_\odot$ for the SMC case ($Z=0.002$), compared with the standard tracks at the end of the model calculation with the self-consistent recipe (second dot of the plots).
	These differences in masses are dimmed later, where we observe that the Galactic case suffer a sudden loss of mass (seen also in the eruptive increment in $\dot M$ from Fig.~\ref{mdot_fin}) before the end of the hydrogen-burning stage and then it finishes with a stellar mass only $\sim4-5$ $M_\odot$ more than the classical track (see Table~\ref{timescalesXY}).
	A similar sudden loss of mass is also evidenced for the LMC case, making the final $M_*$ at the end of the H-burning stage almost similar to the classical track.
	These `eruptive processes' are consequence of the switch in mass-loss rate due to the bi-stability jump around $T_\text{eff}\sim25$ kK \citep{vink99}, according to the Vink's formula.
	\citet{vink99} have suggested that such increase might be related to typical variations shown by Luminous Blue Variable (LBV) stars.
	In addition, \citet{groh14} demonstrated that stars born with more than 60 $M_*$ reach the LBV spectroscopic phase before the ending of the H-core burning stage.
	Recently however, new wind models puts some doubts about the reality of this wind enhancement \citep{bjorklund22}.
	
	Galactic case presents similar stellar radii for both classical and self-consistent tracks at the end of the H-burning stage, despite the radii for the self-consistent track presents an extra peak around $\sim2$ Myr which coincides with the eruptive increase in mass-loss rate.
	This extra peak is also observed for the self-consistent track for the LMC, although here final radii is reduced its $R_*$ a $\sim26\%$ with respect to the classical track.
	However, the most remarkable difference in the stellar radii comes from the SMC case, which generates the largest stellar sizes at the end of the H-core burning stage (see Table~\ref{timescalesXY}) and where final $R_*$ for the self-consistent track is $\sim30\%$ for the classical track.
	This can be attributed to, the eruptive increment in mass-loss rate due to eruptions is not so abrupt compared with the more metallic cases (indeed, this is around one order of magnitude smaller for SMC, as shown in Fig.~\ref{mdot_fin}) and therefore the increasing on stellar radius is not interrupted.
	Such hypothesis is reinforced if we think that LBV stage is not thought to be independent of the mentioned bi-stability jump \citep{vink99}, which is produced when iron recombines from Fe IV to Fe III.
	Naturally, contribution of Fe lines to the line-acceleration in tracks with metallicity $Z=0.002$ is less relevant than the other cases and hence switch in mass-loss rate is smaller.

	Paraphrasing, eruptive processes associated to LBV stage affects the evolution for high metallicities (Galactic and LMC) with different consequences.
	At solar metallicity, stellar radii is almost the same (for classical and self-consistent tracks) and the difference is evidenced in the final mass at the end of the H-burning stage.
	On the contrary, final masses are quite similar for the LMC metallicity but the self-consistent track generates a much smaller star at the end of the H-burning stage.
	According to what is observed for the models with $M_\text{ZAMS}=70$ $M_\odot$ (see following Subsection), we certify the case of similar radii and different final masses is exceptional, and thus it can be attributed to the peculiar conditions of $M_\text{ZAMS}=120$ $M_\odot$ at solar metallicity.
	After all, this is the track with the most extreme mass-loss rates and the only one with remarkable modifications in its surface abundances.
	
	Moreover, these initial cases with $M_\text{ZAMS}=120$ $M_\odot$ are the only instances where the surface abundances have a significant modification at the end of the H-burning stage for all the metallicities (last three columns of Table~\ref{timescalesXY}).
	Self-consistent tracks present slightly more helium at their surfaces at the end of the H-burning stages, particularly for Galactic and LMC cases, even when timescales were marginally shorter.
	We interpret these differences as a consequence of the bigger convective core as shown in Fig.~\ref{ccmass_fin} and of the mass losses.
	Chemical composition is built by the decrease in mass of the convective core. Depending then on the mass loss rate, layers having belonged to those regions partly processed by H-burning are exposed to the surface.
	The value of He enrichment obtained at the surface is thus an interplay between mass-loss and convective core evolution.
	
\subsection{Cases $M_\text{ZAMS}=70\;M_\odot$}
	Here we observe a similar trend compared with the former $M_\text{ZAMS}=120$ $M_\odot$ case, but in a minor scale.
	Tracks using $\dot M_\text{sc}$ retain $\sim15$ $M_\odot$ of extra mass for the Galactic case this time, $\sim10$ $M_\odot$ for the LMC case and $\sim3$ $M_\odot$ for the SMC case, at the end of the self-consistent recipe.
	Again, eruptive increments on mass-loss rate after the end of the stage using $\dot M_\text{sc}$, compensate these initial differences in the evolution of the stellar masses, and once more we obtain close similar $M_*$ at the end of the H-burning stage.
	However, now we observe for all our metallicities that self-consistent tracks predicts smaller stellar radii, although the biggest differences are given for Galactic and LMC cases anew.
	
	And also similar with the $M_\text{ZAMS}=120$ $M_\odot$ case, we observe an extra peak for the stellar radii when solar metallicity is adopted, coinciding with the eruptive increment in $\dot M$.
	Hence, we could argue that the higher retention of mass for the self-consistent tracks produce larger stars that reaching closer to the Eddington limit and then generating larger eruptions in the LBV stage.

\subsection{Cases $M_\text{ZAMS}=40\;M_\odot$ and $M_\text{ZAMS}=25\;M_\odot$}
	For these two initial masses, differences between classical and self-consistent tracks are just marginal, with the expectable increase in luminosity for the HR diagrams (Fig.~\ref{HDR_final}) and slightly larger stellar radii and masses at the end of H-burning stage (Table~\ref{timescalesXY}) even after shorter timescales.
	However, despite these dissimilarities are minor, they reinforce the big trend observed: because $\dot M_\text{sc}<\dot M_\text{Vink}$, self-consistent tracks produce larger and brighter stars, reaching to the end of the H-burning stage in a shorter lifetime.
	
	One particular model deserving attention is the tracks with $M_\text{ZAMS}=25$ and $40$ $M_\odot$ and $Z=0.002$, which reach the end of the H-burning stage without leaving the range of validity for $M_\text{sc}$ and therefore they present only two black dots.
	We can observe such peculiarity in the Fig.~\ref{sHDR_final}, where the track remains only limited to the region $T_\text{eff}\ge32$ kK and $\log g\ge3.2$.
	This is interesting, because it means that m-CAK self-consistent prescription could describe the wind hydrodynamics for massive stars at He-burning stages, when masses are around the $\sim40$ $M_\odot$ and smaller, and metallicities are very low.
	We leave this for a forthcoming study, where we will explore self-consistent solutions for stages with burning processes beyond hydrogen.

\section{Some consequences of the new hydrodynamic wind model}\label{structure}
	\begin{figure}[t!]
		\centering
		\includegraphics[width=0.9\linewidth]{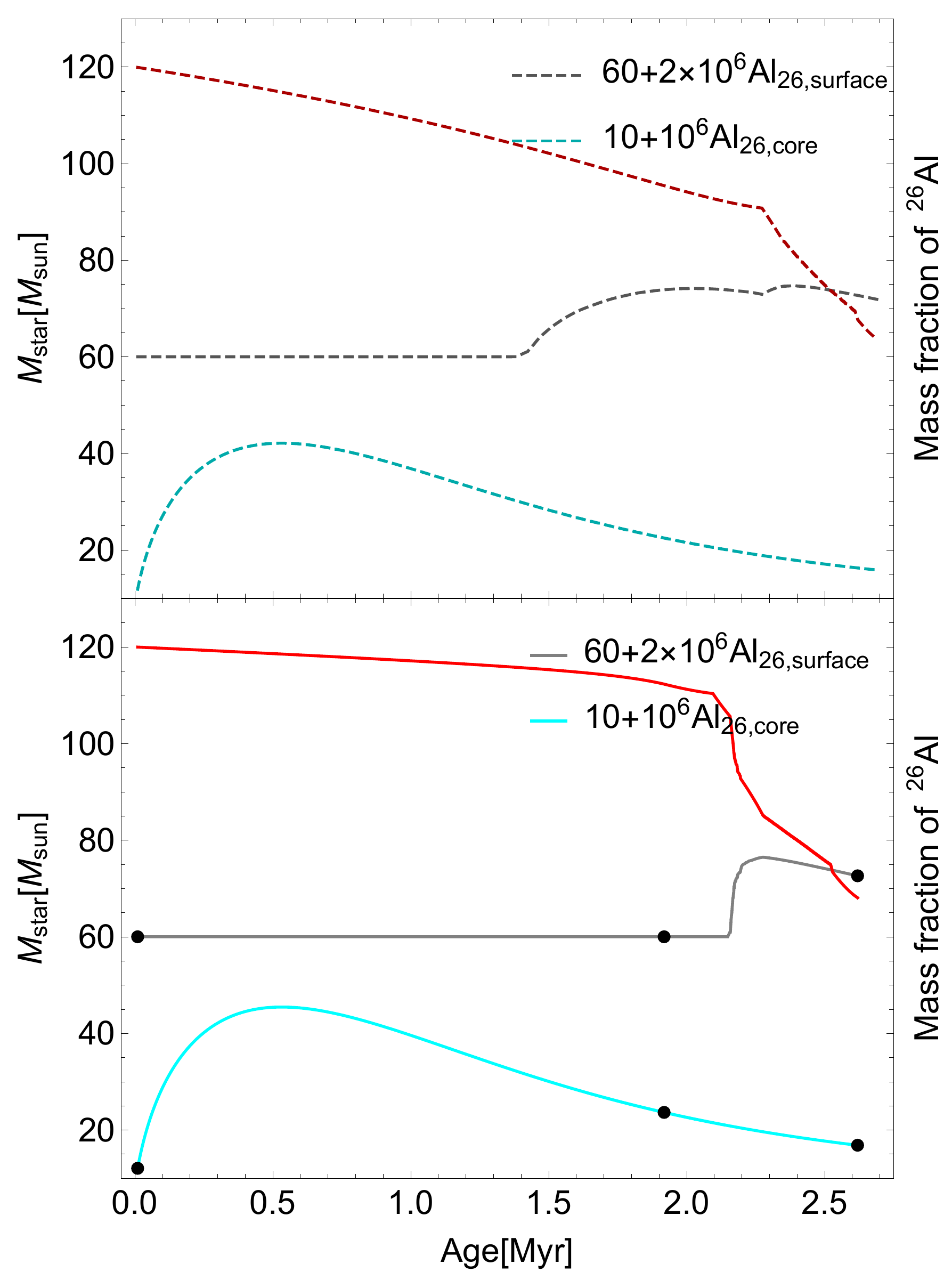}
		\caption{\small{Evolution as a function of time of the total mass (red curve), of the mass fraction of $^{26}$Al at the centre and at the surface of 120 $M_\odot$ at $Z=0.014$.
		The top panel shows the case when the Vink’s formula for the mass-loss rate is adopted.
		The bottom panel shows the case when the new rates obtained from hydrodynamic wind models are used.
		Note that making the plot more easily readable the curves for the mass fraction have been obtained by multiplying the mass fraction by a constant factor and the starting point of each curves has been shifted upwards along the Y axis (see the figure for the adopted values).
		Black dots represent the same as indicated in Fig.~\ref{HDR_final}.}}
		\label{al26plot}
	\end{figure}

	The consequences of the new hydrodynamic wind models remain modest on many outputs of the stellar models as the evolutionary tracks in the HR diagram (Fig.~\ref{HDR_final}) or the evolution of the convective cores (Fig.~\ref{ccmass_fin}).
	On the other hand, for the most massive stars at solar metallicity, the impact on the evolution of the total mass as a function of time is significant (see the left panel of Fig.~\ref{mass_fin} and especially the case for the 120 $M_\odot$).
	Thus in this metallicity/mass domain the adoption of the new hydrodynamic wind models may indeed have a significant impact on some models outputs.

	One obvious consequence is the impact on the wind-enrichment of the interstellar medium in isotopes produced in massive stars during the core-hydrogen burning phase.
	The radionuclide isotope $^{26}$Al is a good example.
	It is produced by proton capture on $^{25}$Mg in the convective cores of stars with masses above about 25 $M_\odot$ \citep[see][]{prantzos86,meynet97,palacios05} and can be ejected by stellar winds.
	Figure~\ref{al26plot} shows how the mass fraction of $^{26}$Al evolves as a function of time at the centre and at the surface of two 120 $M_\odot$ models at solar metallicity computed with the Vink’s formula (top panel) and with the new hydrodynamic wind model (bottom panel).
	We see that in the model computed with the new mass-loss rates, slightly larger (around $\sim10\%$ more) central abundances are reached for $^{26}$Al.
	This is due to the fact that the actual mass remains larger since the mass-loss rates are decreased.
	This allows the convective core to be a bit larger (see the left panel of Fig.~\ref{ccmass_fin}) increasing thus the central temperature and thus allowing larger abundances of $^{26}$Al to be reached (the peak abundance is reached when the production rate by proton capture on $^{25}$Mg is balanced by the decay rate; if the temperatures are larger the proton capture rate is larger and larger abundances are reached).
	However to obtain larger abundances at the centre does not necessarily imply to eject more $^{26}$Al by stellar winds.
	Actually lower winds imply that the layers having once belonged to the core and enriched in $^{26}$Al appear at later stages, when enough time has elapsed for the $^{26}$Al abundance to decay in those layers exposed at the surface.
	This, despite the production of $^{26}$Al in the core is the same when we adopt $\dot M_\text{sc}$ instead of Vink's formula. 
	A consequence is a decrease of the wind contribution of massive stars in $^{26}$Al integrated through lifetime.
	Let us remind here that thanks to the observed diffuse emission at 1.8 MeV produced by the decay of $^{26}$Al present in the Galactic interstellar medium, one knows that presently our Galaxy contains a quantity of $^{26}$Al between 1.7 and 3.5 $M_\odot$ \citep{knodlseder99,diehl06,wang09}.
	Winds of massive stars might contribute to about 1 $M_\odot$ so to about half or one third of the total content in this isotope of the Milky Way \citep[see e.g.][]{palacios05}.
	The new mass-loss rates would decrease this contribution by about a factor two, which can be easily seen if we integrate the mass fraction of $^{26}$Al in the surface over the lifetime of the stellar model (see Table~\ref{integratedal26}).
	A more detailed study needs however to be done to really assess quantitatively the consequences of the new mass-loss rates for the entire lifetime of massive stars including phases as the WR one.
	What is however obvious is that the wind contribution from single stars will be decreased, leaving thus more room for the supernova contribution and binary systems.

	\begin{table}[t!]
		\centering
		\caption{\small{Values for the integration of the mass fractions $^{26}$Al$_\text{surface}$ (see Fig.~\ref{al26plot}).}}
		\begin{tabular}{ccc}
			\hline\hline
			Track & $\int \dot M\;\text{$^{26}$Al}(t)\,dt$\\
			& $[M_\odot]$\\
			\hline
			Classic & $3.47\times10^{-4}$\\
			Self-consistent & $1.82\times10^{-4}$\\
			\hline
		\end{tabular}
		\label{integratedal26}
	\end{table}

	Other significant consequences of these lower mass-loss rates may come out when considering the evolution of massive star with a surface magnetic field.
	Only $8-10\%$ of the OB stars have a surface magnetic field that can be measured \citep{wade18}.
	For those magnetic massive stars, the surface magnetic field can have important effects for slowing down the star through the process of wind magnetic braking \citep{meynet11,keszthelyi21} or by decreasing the mass that is lost by stellar winds through the process of wind magnetic quenching \citep{owocki16,petit17,georgy17}.
	The importance of these processes depends on how the density of kinetic energy in the wind (that depends on the mass-loss rates) compares to the density of magnetic energy.
	Decreasing the mass-loss rates, everything else kept constant, increases the impact of the magnetic field on the evolution of massive stars.
	As indicated above, dedicated models needs be done in order to quantitatively assess how the new mass-loss rates considered here impact these magnetic effects.
	But we suspect that the effects might be significant.

\section{Summary \& Conclusions}\label{conclusions}
	We have implemented our new mass-loss rates, $\dot M_\text{sc}$, in \textsc{Genec} and computed the evolution of massive stars with different initial masses and metallicities with this new prescription.
	Mass-loss rates used for this study are based in the self-consistent m-CAK prescription for wind solutions performed by \citet{alex19,alex21b}, replacing the classical Vink's formula \citep{vink00,vink01} for our new formula.
	
	To achieve this, we developed a recipe for $\dot M_\text{sc,predicted}$ (Eq.~\ref{mdotformula1}) doing a statistical analysis, establishing that the `intervariable fit' (with stellar parameters crossed with each other) provides a more reduced deviation between prediction and real self-consistent mass-loss rate.
	This intervariable fits also allows us to explore the $\dot M\propto Z^a$ relationship, retrieving that the exponent $a$ depends on the stellar mass: the more massive the star, the smaller the $a$ exponent.
	This may be due to the fact that more massive is a star, nearer we are from the Eddington limit.
	In that case the continuum may become more and more important in contributing to the winds.
	The effect of continuum is less dependant on metallicity since it is does not involve interaction between the radiative field and absorption lines.

	Because global values for mass-loss rates provided by the m-CAK prescription are lower than the values provided by the Vink's formula, self-consistent tracks retain more mass during their evolution.
	As a consequence, they produce larger and more luminous stars, as appreciated in the HR diagrams (Fig.~\ref{HDR_final}).
	Such differences are more prominent for our models with higher initial masses, confirming that mass-loss rate plays a more relevant role over the stellar evolution for very massive stars in the order of $M_*\gtrsim60$ $M_\odot$ \citep{vink12}.
	These larger radii for the self-consistent tracks also makes the evolution model to approach closer to the Eddington limit, and then to experience the eruptive increases in mass-loss rate associated to the Luminous Blue Variable stage prior to the end of the H-burning stage \citep{groh14}.
	Besides, we observe that the most massive stars present slightly shorter lifetime and end their H-burning stage with a larger percentage of helium abundance in their surface.
			
	Nevertheless, the most remarkable difference between classic and self-consistent tracks is the generation and abundance of  $^{26}$Al in the stellar surface of our model with $M_*=120\;M_\odot$ and $Z=0.014$, which suggest that the contribution of this isotope to Galactic interstellar medium coming from stellar winds is less than previously thought to be.
	
	These promising results open the door to the necessity to extend this current study for a wider range of evolutionary stages and the inclusion of rotation, together with the constrain of populations of massive stars in the Galaxy and beyond.
	Because self-consistent tracks are based on smaller mass-loss rates (and subsequently more massive and larger stars), we could speculate that the loss of angular momentum should be lower.
	Such extra elements, and their consequences, will be included in a forthcoming paper.
		
\begin{acknowledgements}
	This project has received funding from the European Unions Framework Programme for Research and Innovation Horizon 2020 (2014-2020) under the Marie Sk{\l}odowska-Curie grant Agreement N$^{\rm o}$ 823734.
	ACGM has been financially supported by the PhD Scholarship folio N$^{\rm o}$ 21161426 from National Commission for Scientific and Technological Research of Chile (CONICYT).
	ACGM and MC acknowledge support from Centro de Astrof\'isica de Valpara\'iso.
	MC thanks the support from FONDECYT project 1190485.
	JC thanks the support from FONDECYT project 1211429.
	ACGM and JC are grateful of the support from Max Planck Partner Group on Galactic Centre Astrophysics.
	GM has received funding from the European Research Council (ERC) under the European Union's Horizon 2020 research and innovation programme (grant agreement No 833925, project STAREX).
	JHG and LM wish to acknowledge the Irish Research Council for funding this research.
\end{acknowledgements}

\bibliography{evolpaper.bib} 
\bibliographystyle{aa} 

\newpage
\begin{appendix}
\section{Alternative fits for $\dot M_\text{sc,predicted}$}\label{extrafittings}
	From Section~\ref{statisticalfitting}, our analysis determined that the most optimal fit to reduce discrepancies between the predicted $\dot M$ from a recipe and the real values for $\dot M$ from our grid of self-consistent solutions (Tables~\ref{standardtable}, \ref{standardtablez07} and \ref{standardtablez03}) is the intervariable fitting.
	Notwithstanding the foregoing, we decide to show in this Appendix the alternative parameters for the best linear and quadratic fittings, whose coefficient of determination $R^2$ is tabulated in Table~\ref{r2table}.

\subsection{Linear fit}
	\begin{align}\label{mdotformula2}
		\log\dot M_\text{sc,predicted}=&-16.686 + 8.167\,w\nonumber\\
		& - 1.281\,x + 2.05\,y + 0.669\,z\;,
	\end{align}
	where $w$, $x$, $y$ and $z$ are defined as:
	\begin{equation}
		w=\log \left(\frac{T_\text{eff}}{\text{kK}}\right)\;,\nonumber\\
		x=\log g\;,\\
		y=\log\left(\frac{R_*}{R_\odot}\right)\;,\\
		z=\log \left(\frac{Z_*}{Z_\odot}\right)\;.
	\end{equation}

\subsection{Quadratic fit}
	\begin{align}\label{mdotformula3}
		\log\dot M_\text{sc,predicted}=&-12.657 + 21.699\,w - 4.808\,w^2\nonumber\\
		&- 10.957\,x + 1.356\,x^2 + 7.899\,y \nonumber\\
		&- 2.212\,y^2 + 0.775\,z + 0.165\,z^2\;,
	\end{align}
	where $w$, $x$, $y$ and $z$ are defined as:
	\begin{equation}
		w=\log \left(\frac{T_\text{eff}}{\text{kK}}\right)\;,\nonumber\\
		x=\log g\;,\\
		y=\log\left(\frac{R_*}{R_\odot}\right)\;,\\
		z=\log \left(\frac{Z_*}{Z_\odot}\right)\;.
	\end{equation}

\end{appendix}

\end{document}